\documentclass[%
 reprint,
superscriptaddress,
 amsmath,amssymb,
 aps,
prb,
]{revtex4-2}

\renewcommand{\thefigure}{\arabic{figure}}
\renewcommand{\figurename}{Fig.}
\usepackage[normalem]{ulem}
\usepackage{xcolor}

\usepackage{multirow}
\usepackage{booktabs}
\usepackage{xcolor}

\usepackage{graphicx}
\usepackage{dcolumn}
\usepackage{bm}

\begin{document}

\title{Order–Disorder in Fe–Si Alloys: Implications for Seismic and Transport Properties of Earth's Inner Core}

\author{Cong Liu}
\email{Corresponding author: cliu10@carnegiescience.edu}
\affiliation{Extreme Materials Initiative, Earth and Planets Laboratory, Carnegie Institution for Science, 5241 Broad Branch Road NW, Washington, District of Columbia 20015, USA}
\author{Xin Deng}
\affiliation{Extreme Materials Initiative, Earth and Planets Laboratory, Carnegie Institution for Science, 5241 Broad Branch Road NW, Washington, District of Columbia 20015, USA}
\author{R. E. Cohen}
\email{rcohen@carnegiescience.edu}
\affiliation{Extreme Materials Initiative, Earth and Planets Laboratory, Carnegie Institution for Science, 5241 Broad Branch Road NW, Washington, District of Columbia 20015, USA}
\date{\today}

\begin{abstract}

Understanding the structure and dynamics of Earth's inner core is essential
for constraining its composition, thermal evolution, and seismic behavior.
Silicon is widely considered a major light element in the core. Here, we use
first-principles molecular dynamics and thermodynamic modeling to investigate
the structural, elastic, and transport properties of iron-silicon (Fe-Si)
alloys under inner-core conditions. By computing the Gibbs free energies of
multiple solid phases, including ordered B2, hexagonal close-packed (hcp),
face-centered cubic (fcc), and body-centered cubic (bcc) Fe-Si, we construct a
phase diagram relevant to inner-core pressures and temperatures. We identify a
pronounced miscibility gap between hcp and cubic Fe--Si phases, with phase
coexistence over approximately 6--11 wt\% Si at 6000 K. The cubic phase exhibits
a strong order-dependent elastic response, with disordered bcc showing a large intrinsic shear anisotropy (22.9\% at 6000 K), in contrast to the nearly
isotropic hcp phase (0.6\%). Meanwhile, disordered bcc Fe--Si produces shear
wave velocities ($V_S=3.73$ km$\cdot$s$^{-1}$) and Poisson's ratios comparable
with seismological constraints. In addition, disordered bcc Fe--Si exhibits a
lower thermal conductivity than hcp Fe--Si, providing a quantitative constraint
on phase-dependent electronic heat transport. These results demonstrate that
chemical ordering strongly controls the stability and physical properties of
cubic Fe--Si alloys, and suggest that metastable or localized cubic Fe--Si
phases may contribute to the structural and transport complexity of Earth's
inner core.

\end{abstract}

\maketitle



\section*{Introduction}
Whether Earth’s solid inner core adopts the hexagonal close-packed (hcp) or body-centered cubic (bcc) structure of iron has remained a subject of long-standing debate \cite{brownEquationStateIron2001,dubrovinskyBodyCenteredCubicIronNickel2007a,mikhaylushkinPureIronCompressed2007,tatenoStructureIronEarths2010,anzelliniMeltingIronEarths2013,krausMeasuringMeltingCurve2022a}. Static and dynamic compression experiments show that pure Fe remains hcp structured through the Earth's inner core pressures and temperatures. \cite{tatenoStructureIronEarths2010,krausMeasuringMeltingCurve2022a}. Ab initio molecular-dynamics free-energy calculations show that although pure bcc Fe is entropically favored at core temperatures, it remains higher free energy than hcp Fe \cite{vocadloPossibleThermalChemical2003,belonoshkoStabilityBodycentredcubicPhase2003,luoDynamicalStabilityBody2010a,godwalStabilityIronCrystal2015,sunTwostepNucleationEarths2022a,sunUnveilingEffectNi2024a}. 
Seismic observations indicate a finite shear-wave anisotropy of the inner core, with the observed S-wave velocity difference between different propagation directions being approximately 0.8--1\% \cite{wangShearWaveAnisotropyEarths2021}. However, hcp Fe is predicted to exhibit nearly isotropic sound velocities under inner-core conditions \cite{shaElasticIsotropyEFe2010}, providing a weaker intrinsic anisotropic contribution compared with bcc Fe \cite{Sun2008,Frost2019,Wu2025,Vidale2025}. The observed anisotropy is also much smaller than the intrinsic single-crystal elastic anisotropy predicted for bcc Fe, which can reach 15--20\% depending on the elastic anisotropy metric and temperature conditions\cite{steinle-neumannElasticityIronTemperature2001a,vocadloInitioCalculationsElasticity2007,belonoshkoOriginLowRigidity2007,tkalcicShearPropertiesEarths2018,belonoshkoElasticPropertiesBodycentered2022,ikutaSoundVelocityHexagonal2022,costadelimaEstimateAbsoluteShearwave2023}. The aggregate average seismic anisotropy averaged over crystals cannot be larger than that of the intrinsic crystal anisotropy, so the inner core cannot be pure hcp iron alloy.

Light elements such as H, C, O, Si, and S, incorporated during planetary accretion and differentiation \cite{li315ExperimentalConstraints2014,hiroseLightElementsEarths2021a} are inferred to reside in both the liquid outer core, which contains up to 20 wt.\% light elements, and the solid inner core (IC), with a much smaller fraction \cite{hiroseLightElementsEarths2021a}. First-principles calculations and high-pressure experiments have long been employed to constrain the composition of Earth’s core \cite{Stixrude1997,alfeCompositionTemperatureEarths2002}. Among the potential light elements, silicon stands out as a particularly compelling candidate \cite{Fischer2015, Rubie2015}. Geochemical evidence also suggests that silicon is concentrated in the core; Si isotope systematics in terrestrial and lunar basalts reveal a heavy $\delta^{30}$Si signature consistent with early segregation of silicon into Earth's metallic core \cite{georgSiliconEarthsCore2007}. Estimates of silicon content in the solid inner core vary, with recent studies suggesting approximately 4.1 wt.\% Si \cite{liuHydrogenSiliconAre2024}, whereas others propose lower values around 2.3 wt.\% \cite{hiroseLightElementsEarths2021a}. These studies, however, generally assume that Si is incorporated into hcp Fe alloys within the inner core. High-pressure experiments on Fe–Si alloys \cite{dubrovinskyIronSilicaInteraction2003, linPhaseRelationsFeSi2009, fischerPhaseRelationsFe2013a, tatenoStructureFeSi2015, ikutaTwophaseMixtureIron2021b,edmundFeFeSiPhaseDiagram2022a, yokooCompositiondependentThermalEquation2023a, nagayaEquationsStateB22023} 
demonstrate that silicon expands the stability field of bcc/B2 phases near the melting curve, and that cubic and hcp structures can coexist over certain composition ranges \cite{fuCoreOriginSeismic2023a}. The coexistence of these phases in the inner core allows higher Si concentrations, consistent with seismological constraints, as demonstrated below. Computational studies \cite{coteInitioLatticeDynamics2010,cuiEffectSiStability2013,liCompetingPhasesIron2024a,liShortrangeOrderStabilizes2025} highlight the entropy-driven competition among hcp, fcc, bcc, and B2 structures under the high temperatures of Earth's core. 

\begin{figure*}[htp]
    \centering
    \includegraphics[width=0.9\linewidth]{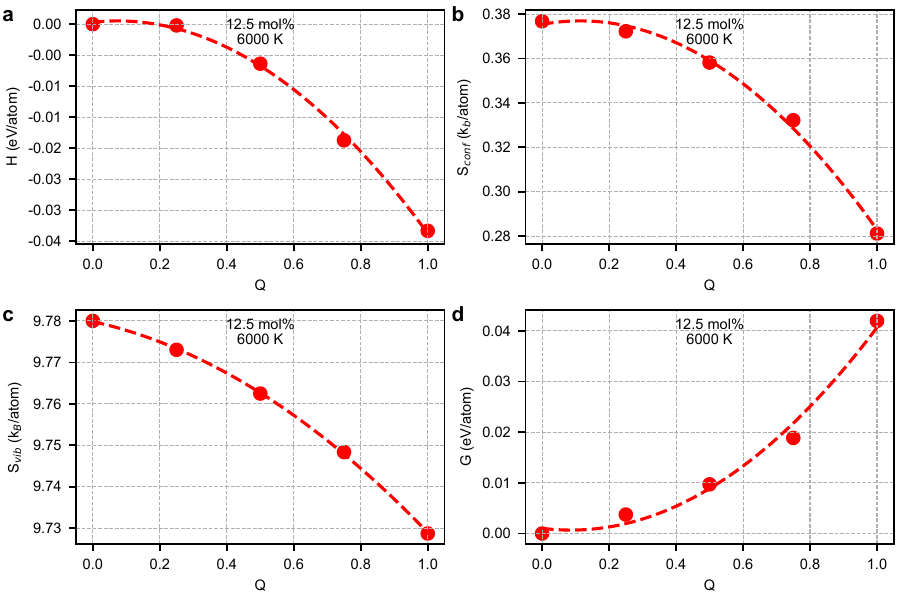}
    \caption{\textbf{Effects of the order parameter in B2 Fe–Si alloys} illustrated for a representative composition of Fe$_{0.875}$Si$_{0.125}$ (12.5 mol\% Si), where $Q=0$ and $Q=1$ denote completely disordered (bcc) and fully ordered (B2) states, respectively. \textbf{a} Enthalpy at 330 GPa 6000 K, \textbf{b} configurational entropy, \textbf{c} vibrational entropy at 330 GPa and 6000 K, and \textbf{d} Gibbs free energy at 330 GPa and 6000 K as functions of order parameter $Q$. The enthalpies and Gibbs free energy at Q=0 have been shifted to zero.}
    \label{fig:bcc}
\end{figure*}

Whereas pure iron crystallizes in the hcp phase under inner-core conditions \cite{krausMeasuringMeltingCurve2022a}, the pure compound FeSi adopts the B2 (CsCl-type) structure \cite{fischerPhaseRelationsFe2013a} at low to moderate temperatures. For solid solutions of Fe with FeSi, the excess Fe partitions between Fe and Si sites within the B2 lattice. This represents a case of long-range convergent ordering \cite{Golosov1975, Holland1996, Jiang2004, Kiyokane2017}. If the Fe and Si atoms are fully disordered, the two sites become equivalent, producing the bcc structure; partial ordering yields the B2 phase. Here we adopt a similar approach to that recently applied to high-pressure Mg$_2$GeO$_4$, in which Mg and Ge disorder on different crystallographic sites, i.e., non-convergently \cite{duttaUltrahighpressureDisorderedEightcoordinated2022,duttaHighpressureOrderdisorderTransition2023, Zheng_2025}. 

Fe--Si alloys exhibit variable degrees of chemical order depending on $P$, $T$, and composition. Theoretical studies have considered both ideal and non-ideal \cite{liShortrangeOrderStabilizes2025} mixing at high $P$--$T$ conditions. Here we combine density functional theory calculations, the special quasirandom structure (SQS) method \cite{Zunger1990,vandeWalle2013}, and \textit{ab initio} molecular dynamics and compute the Fe--Si binary phase diagram and quantify how order--disorder transitions influence thermal and electrical conductivity as well as seismic anisotropy of Fe--Si alloys under Earth’s inner-core conditions.

\section*{Results}

We constructed long-range ordered configurations of B2, hcp, and fcc structures for Fe$_{1-X}$Si$_X$ for mole fraction $X$ from 0 to 0.5, varying the Si occupancy at each inequivalent lattice site (Fig.~S1 in the Supplemental Materials). For hcp and fcc, we also allow partitioning of Si atoms across two inequivalent sites, breaking the hcp and fcc symmetry. Each Fe–Si composition and site concentration (Table~S1 in the supplemental materials) were used to generate quasi-random configurations in a 4$\times$4$\times$4 (128-atom) supercell with an order parameter $Q$ ranging from 0 (fully disordered) to 1 (fully ordered) (see Methods). Enthalpies of these configurations were obtained from first-principles molecular dynamics (FPMD) simulations at selected temperatures, averaged over equilibrium segments of the trajectory. As shown in Fig.~S2 in the Supplemental Materials, the enthalpy of the B2 phase decreases significantly with increasing $Q$, particularly at higher Si concentrations. At zero temperature, pure hcp Fe coexists with pure ordered FeSi.

\begin{figure*}[htb]
    \centering
    \includegraphics[width=0.9\linewidth]{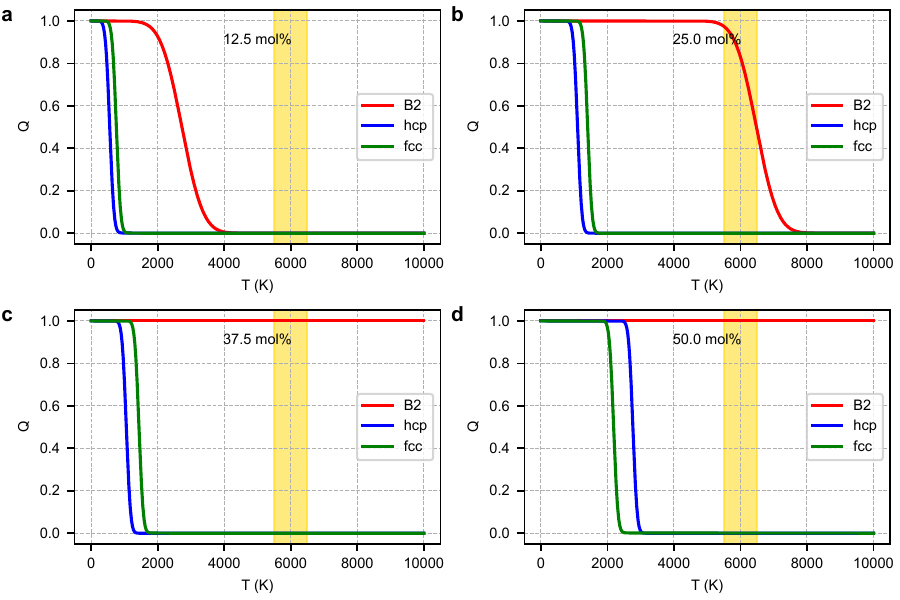}
    \caption{\textbf{Order–disorder phase transition in Fe–Si alloys.} Temperature dependence of the order parameter ($Q$) for the B2 (red), hcp (blue), and fcc (green) phases of Fe–Si alloys at Si content of \textbf{a} 12.5, \textbf{b} 25.0, \textbf{c} 37.5, and \textbf{d} 50 mol\%, corresponding to Fe$_{0.875}$Si$_{0.125}$, Fe$_{0.75}$Si$_{0.25}$, Fe$_{0.625}$Si$_{0.375}$, and Fe$_{0.5}$Si$_{0.5}$, respectively. The yellow shaded region indicates the estimated temperature range of Earth’s inner core \cite{anzelliniMeltingIronEarths2013}. The disordered phase ($Q=0$) of the B2 phase corresponds to the bcc structure.}
    \label{fig:Q}
\end{figure*}

The Gibbs free energy as a function of the order parameter ($Q$) can be expressed in terms of enthalpy, configurational entropy, and vibrational entropy:
\begin{align}
    G(P,T) &= H(T, V) - (S_{\text{conf}} + S_{\text{vib}})T, \\
           &= E(T, V) + PV + ZPE - (S_{\text{conf}} + S_{\text{vib}})T,
\end{align}
where $E(T, V)$ is the total internal energy, including vibrational contributions; $ZPE$ is the zero-point energy; and $S_{\text{vib}}$ is the vibrational entropy, all obtained from FPMD simulations at fixed temperature. Different supercell configurations corresponding to $Q$ values between 0 and 1 were generated to capture the full order–disorder phase relationship (see Methods). The enthalpy of the B2 structure (Fig.~\ref{fig:bcc}\textbf{a}) indicates that at low temperature Si atoms preferentially occupy one of the two sublattice sites in the two-site model ($Q=1$), rather than being equally distributed between the two sites as in the disordered state ($Q=0$). The disordered configuration exhibits substantially higher total entropy, with approximate contributions of 0.10~$k_B$ from configurational and 0.05~$k_B$ from vibrational sources. Consequently, the Gibbs free energy curves (Fig.~\ref{fig:bcc}\textbf{d}) indicate that Fe$_{0.875}$Si$_{0.125}$ favors the disordered bcc state over the ordered B2 phase by $\sim$41~meV per atom at 330~GPa and 6000~K.

The competition between configurational entropy and interatomic interactions governs the degree of Fe-Si ordering. Configurational entropy favors complete disorder, stabilizing the bcc phase in which Si atoms randomly occupy both bcc sublattices with equal probability, consistent with the ideal mixing approximation \cite{alfeConstraintsCompositionEarths2000,alfeCompositionTemperatureEarths2002}. This entropic effect becomes increasingly dominant at higher temperatures, driving an order–to-disorder transition that is either continuous or discontinuous depending on the curvature of the Gibbs free energy surface with respect to $Q$, $P$, $T$, and $X$. The vibrational entropy, S$_{vib}$,  also depends on Q, so we also computed the vibrational entropy versus Q. 

By identifying the local minima of the Gibbs free energy at varying temperatures, we obtained the temperature dependence of the order parameter ($Q$) for the cubic (B2, bcc), hcp, and fcc phases of Fe–Si alloys with different compositions (Fig.~\ref{fig:Q}). At 12.5~mol\% Si, the cubic, hcp, and fcc phases all become disordered under inner-core temperatures. Specifically, the B2 phase transforms into a fully disordered bcc structure above $\sim$4000~K at 330~GPa,(Fig.~\ref{fig:bcc}) and the hcp and fcc phases undergo disordering at much lower temperatures, near 1000~K. The transition temperature systematically increases with Si content; the B2 phase remains ordered even at 6000~K for Si concentrations exceeding 25~mol\%. In contrast, both hcp and fcc Fe–Si alloys remain disordered across the entire composition range (0–50~mol\%) at 330~GPa and 6000~K.

\begin{figure*}[htp]
    \centering
    \includegraphics[width=0.7\linewidth]{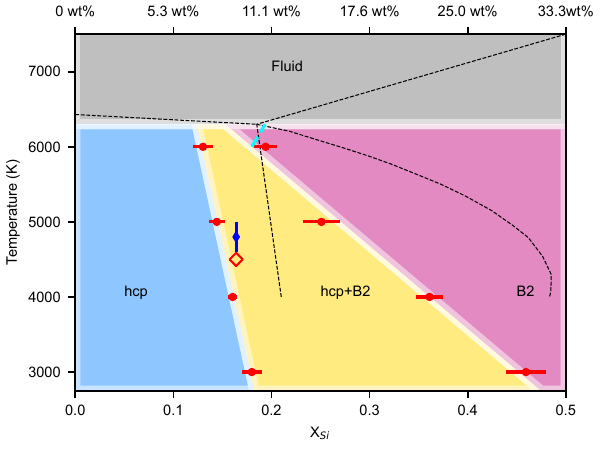}
    \caption{\textbf{Calculated solid-state Fe–Si phase diagram at 330~GPa.} Red points denote phase boundaries separating hcp (blue), coexisting hcp+B2 (yellow), and B2 (magenta) regions, determined from common-tangent constructions between Gibbs free energy curves of the B2 and hcp phases at different temperatures. The ordered B2–Fe–Si alloy transforms into a disordered bcc phase ($Q = 0$) above the dashed cyan line. The dashed black lines represent the phase boundaries (hcp, coexistence, B2, and fluid) determined in Ref.~\onlinecite{liShortrangeOrderStabilizes2025}. The shaded grey regions indicate the constraints imposed by melting curves beyond the range of our simulations. The blue diamond indicates the experimental constraints \cite{tatenoStructureFeSi2015} defining the boundary between the hcp phase and the hcp/B2 mixture for Fe–9Si at 330 GPa. The open red diamond is extrapolated from lower-pressure experiments \cite{fischerPhaseRelationsFe2013a}.}
    \label{fig:phase}
\end{figure*}

The disordered bcc phase of Fe–Si alloys is stabilized only at high temperatures, whereas the ordered B2 structure has been consistently observed in X-ray diffraction (XRD) experiments \cite{fischerPhaseRelationsFe2013a}. Our calculated XRD patterns for Fe–Si alloy at 330~GPa and 6000~K (Fig.~S3 in the Supplemental Materials) show that the bcc phase is characterized by a set of reflections (e.g., $\langle$110$\rangle$, $\langle$200$\rangle$), and the ordered B2 phase exhibits additional peaks (e.g., $\langle$100$\rangle$, $\langle$111$\rangle$, and $\langle$210$\rangle$). According to Bragg’s law ($n \lambda = 2d\sin \theta$), the bcc peaks shift to smaller diffraction angles relative to those of the B2 phase, reflecting the volume expansion accompanying disorder. This shift arises from increased interatomic spacing as Si atoms become randomly distributed in the disordered bcc lattice. The loss of long-range order thus reduces packing efficiency and weakens directional bonding, leading to a modest lattice expansion of the disordered bcc phase.

We constructed Gibbs free energy curves for varying Si composition within the relevant crystalline phases, including B2, hcp, fcc, and bcc (Fig.~S4 in the Supplemental Materials). Naturally, pure Fe lacks a configurational entropy contribution because there is no compositional disorder. Hcp Fe remains the most thermodynamically stable phase across the studied temperature range. With increasing temperature, the free energy difference between the hcp and bcc iron progressively decreases, due to the higher vibrational entropy of the bcc phase relative to hcp Fe, consistent with previous studies \cite{luoDynamicalStabilityBody2010a,belonoshkoStabilizationBodycentredCubic2017c}.

In Fe$_{0.5}$Si$_{0.5}$, the ordered B2 phase is thermodynamically more stable than the hcp and fcc structures at low temperatures, primarily due to its lower internal energy (Fig.~S4 in the Supplemental Materials). However, the B2 phase exhibits substantially lower configurational and vibrational entropy than the disordered bcc solid solution. Although the hcp phase is energetically less favorable at low temperatures, it has greater atomic disorder and, correspondingly, larger entropy contributions. As temperature increases, these entropic effects progressively narrow the free-energy gap between the ordered B2 phase and other disordered phases, such as the hcp phase.

At intermediate Si contents (0–50~mol\%), the Gibbs free energy landscape is strongly influenced by the order–disorder transition within the B2 structure. For Si concentrations below 12.5~mol\%, the Gibbs free energies of the B2 and bcc phases converge, indicating an order–disorder transition from B2 to bcc.  The free energy of the B2 Fe$_{0.875}$Si$_{0.125}$ becomes comparable to that of the hcp and fcc phases above 5000~K, highlighting the thermodynamic competition among these structures at elevated temperatures. In contrast, for Si content exceeding 25~mol\%, no clear order–disorder transition is observed in the B2 phase up to 6000~K (Fig.~\ref{fig:Q}). The fully ordered B2 phase ($Q=1$) exhibits substantially lower configurational and vibrational entropies (Fig.~S5 and S6 in the Supplemental Materials) than the hcp and fcc phases, which remain disordered ($Q=0$). Consequently, the free energy difference between the B2 and hcp phases decreases with increasing temperature.

Using the composition-dependent Gibbs free energy curves, we constructed the Fe–Si binary phase diagram using common tangent constructions between competing phases (Fig.~\ref{fig:phase}). At each temperature, the two common tangent points define the boundaries separating single- and two-phase regions. The resulting phase diagram reveals three distinct compositional regimes: a low-Si hcp phase, an intermediate two-phase field comprising hcp and cubic Fe--Si phases with varying degrees of chemical order, and a high-Si region dominated by ordered B2 phases. At 3000~K, the Fe–Si system exhibits a pronounced miscibility gap between the hcp and B2 phases, allowing their coexistence across a broad compositional range of approximately 16–46~mol\%~Si. Both boundaries of the two-phase region show negative slopes with respect to temperature, reflecting the higher total entropy of the B2 and bcc phases compared to the hcp phase, consistent with Fe–Si and Fe–Ni–Si alloy experiments at lower pressures \cite{fischerPhaseRelationsFe2013a,ikutaTwophaseMixtureIron2021b}. The steep, low-Si boundary coincides with the experimentally determined Fe–9 wt\% Si phase boundary at 330 GPa \cite{tatenoStructureFeSi2015}, in contrast to the more gradual evolution of the high-Si boundary. This asymmetry arises from differences in entropy and enthalpy contributions across the composition range: at low Si contents, the disordered B2 phase ($Q=0$, bcc) possesses higher entropy and enthalpy than the hcp phase, whereas at high Si contents, the ordered B2 phase ($Q=1$, B2) exhibits lower entropy and enthalpy. These opposing trends result in a progressive narrowing of the miscibility gap as the temperature increases.

The low-Si boundary of the miscibility gap is well described by a linear fit, whereas the high-Si boundary exhibits larger uncertainty. This asymmetry arises from both numerical and thermodynamic factors. At low Si contents, the Gibbs free energy curve of the hcp phase displays large curvature (Fig.~S4 in the Supplemental Materials), making the tangent point well defined and relatively insensitive to small compositional fluctuations. In contrast, the Gibbs free energy surface of the high-Si B2 phase is much flatter due to its lower configurational entropy and weaker compositional dependence, resulting in smaller curvature and, consequently, greater uncertainty in the tangent construction. Because the positions of the two tangent points are coupled through the common tangent condition, any minor deviation in the low-Si region propagates to the high-Si boundary, amplifying its uncertainty. This effect is further enhanced by the finite configurational sampling of the 128-atom supercell used in this study. 
 This entropy difference raises the melting temperature due to the order--disorder transition.

Furthermore, the core temperature is influenced by the melting behavior of Fe--Si alloys. In the case of disordered bcc Fe$_{0.875}$Si$_{0.125}$ (\(Q = 0\)), both vibrational and configurational entropies (Fig.~\ref{fig:bcc}b and c) are elevated by 0.05~\(k_{\mathrm{B}}\) and 0.10~\(k_{\mathrm{B}}\), respectively, relative to the fully ordered B2 phase (\(Q = 1\)). This increase in entropy implies that the order--disorder transition raises the melting temperature of the Fe--Si alloy compared to the ordered configurations typically assumed in earlier melting studies~\cite{belonoshkoMeltingFeFe09375Si006252009}. Consequently, this effect may play an important role in shaping the thermal evolution and crystallization history of Earth’s early core.

\section*{Discussion}

\subsection*{Phase Diagram Comparison}

We observe several key differences between our phase diagram and that of
Ref.~\onlinecite{liShortrangeOrderStabilizes2025}, who employed hybrid Monte
Carlo simulations with deep-learning interatomic potentials. Firstly, our
results reveal a B2--hcp two-phase coexistence field that extends to lower Si
concentrations (approximately 12--16 mol\%) across the 3000--6000 K range. In
contrast, their phase diagram locates the corresponding hcp--B2 boundary at
higher Si content ($\sim$20 mol\%). Secondly, our results differ in the nature of the bcc--B2 transition.
Ref.~\onlinecite{liShortrangeOrderStabilizes2025} reported a persistent
miscibility gap between a short-range-ordered bcc phase and a long-range-ordered
B2 phase. This gap expands upon cooling and remains open even at 0~K,
indicating a first-order transition. In contrast, our calculations reveal no
such coexistence region: the bcc phase transforms continuously into the B2
phase at low Si concentration (Fig.~S7). The Gibbs free-energy curves for bcc
and B2 merge smoothly over the entire 3000--6000~K temperature range
(Fig.~S4), suggesting a second-order (continuous) transformation. Furthermore,
the bcc--B2 boundary in our study occurs at lower Si concentrations than in
their phase diagram, resulting in a narrower miscibility gap. Thirdly, analysis of the chemical potential (Fig.~S8 and Eq.~4 in Supplemental Information) further supports this continuous transition. The chemical potential curves for B2 and bcc phases merge smoothly at low Si concentrations
($\sim$12.5\%), in contrast to the discontinuous behavior associated with a
first-order transition reported in Ref.~\onlinecite{liShortrangeOrderStabilizes2025}.
Finally, compared to Ref.~\onlinecite{liShortrangeOrderStabilizes2025}, we
identify a narrower miscibility gap, shifted toward lower Si concentrations,
and a distinct disordered bcc stability pocket emerging above 5100 K, whereas the cubic
stability region in their calculations appears at higher Si concentration and
temperature. The differences maybe due to two issues in the previous work\cite{liCompetingPhasesIron2024a,liShortrangeOrderStabilizes2025}. Firstly, it seems they did termodynamic integration to determine the Helmholtz free energy in a constant pressure rather than constant volume ensemble, and secondly that the PV term was neglected from the Gibbs Free energy; we find close agreement with their work if we delete the necessary PV term. The general agreement of our results with theirs shows the stability of the predictions.

Our results are broadly consistent with experimental constraints on Fe--Si
alloys at high pressure
\cite{fischerPhaseRelationsFe2013a,tatenoStructureFeSi2015},
but the experiments do not resolve the detailed slope of the hcp--B2 phase
boundary because only a single Si concentration (Fe--9 wt\% Si, i.e.,
16 mol\%) has been measured at inner-core pressures. Fischer \textit{et al.}
mapped the Fe--Si phase diagram only up to 145 GPa, and thus cannot provide
direct constraints at 330 GPa. By extrapolation to 330 GPa, we find that for
Fe--9 wt\% Si the B2 phase becomes competitive with hcp above approximately
4500 K, consistent with the experimental observation that Fe--9 wt\% Si
remains in the hcp structure up to approximately 4800 K before transforming
\cite{tatenoStructureFeSi2015}.

The present phase diagram describes the relative stability of solid Fe--Si
phases. The high-temperature solid boundaries are compared with available
melting constraints, while a fully self-consistent solid--liquid phase diagram
would require dedicated melting calculations.

\subsection*{Cubic phase stability and inner-core stratification}

Although silicon has long been recognized as a primary light element in Earth's
core \cite{georgSiliconEarthsCore2007}, recent estimates based on the stability
of the hcp phase
\cite{hiroseLightElementsEarths2021a,liuHydrogenSiliconAre2024}
suggest an abundance of approximately 2.3--4.1~wt.\% (4.5--7.9~mol.\%).
According to our composition--temperature phase diagram, disordered bcc
Fe--Si is not the equilibrium phase at these plausible inner-core Si
concentrations, implying that hcp should remain the predominant solid phase in
a fully equilibrated system.

During inner-core crystallization, however, the bcc phase may play a role in
nucleation. Yang Sun \textit{et al.}\cite{sunTwostepNucleationEarths2022a}
showed that metastable bcc iron can nucleate more readily than hcp under
inner-core conditions. In this two-step mechanism, bcc forms first and
subsequently transforms into hcp, reducing the undercooling required for
solidification.

Furthermore, experimental evidence
\cite{fuCoreOriginSeismic2023a} indicates that hydrogen may enhance the silicon
concentration in B2 crystals, increasing the Si/Fe molar ratio to approximately
1, and that B2 crystals may coexist with Fe liquid under outer-core conditions.
Such chemical environments may therefore affect light-element partitioning
during core crystallization. To assess the stability of these compositions,
we compared the Gibbs free energy of a mechanical mixture of pure B2 FeSi and
pure hcp Fe with that of coexisting hcp and cubic Fe--Si alloys
(Fig.~S4 in the Supplemental Materials). At 3000~K, the free energy of the
FeSi + Fe mixture is only slightly higher than that of the hcp + cubic alloy
mixture, and this difference increases with temperature, indicating that pure
B2 FeSi is not favored under inner-core conditions.

Within the hcp--cubic two-phase field, the relevant density comparison is
between the two phases at their equilibrium coexistence compositions rather
than at the same composition. At 330~GPa and 6000~K, the common-tangent
construction gives approximately 12.69~mol\% Si for the hcp phase and
20.83~mol\% Si for the cubic phase (Fig.~S14 in the Supplemental Materials).
The corresponding atomic volumes obtained by interpolation of the
volume--composition relations are approximately 6.8799 and 6.8656
\AA$^3$/atom, respectively, yielding densities of approximately
12.63 and 12.11~g\,cm$^{-3}$. Thus, the Si-richer cubic phase is
approximately 4\% less dense than the coexisting Si-poorer hcp phase.

This density contrast suggests a possible tendency for compositional and
phase variations during inner-core evolution. If hcp and cubic Fe--Si
phases coexist, the denser Si-poorer hcp component would be gravitationally
favored at greater depth, whereas the lighter Si-richer cubic component
could become relatively enriched toward shallower regions. However, the
actual spatial distribution depends on compositional partitioning, phase
fractions, pressure, temperature, and crystallization kinetics.

We note that our calculated pure-Fe bcc--hcp Gibbs free-energy separation is
larger than the previous thermodynamic-integration estimate
\cite{liShortrangeOrderStabilizes2025}. As discussed in the Supplemental
Materials, the pressure--volume (PV term) contribution is substantial at inner-core
pressures. Importantly, the larger separation in our calculations disfavors
rather than artificially stabilizes the cubic phase.

These thermodynamic, kinetic, and compositional considerations indicate that
hcp remains the predominant equilibrium phase at plausible inner-core Si
concentrations, whereas cubic Fe–Si may occur as a localized or composition-dependent component during crystallization
and subsequent inner-core evolution. The present calculations therefore do not
imply a dominant cubic inner core or a permanent Si-rich innermost layer.

Because disordered bcc Fe--Si exhibits substantially larger intrinsic
single-crystal elastic anisotropy than hcp Fe--Si, a radial variation in the
fraction of cubic material could contribute to depth-dependent seismic
anisotropy. Quantitative comparison with seismic observations, however,
additionally requires constraints on crystallographic texture, phase
proportions, and their radial and lateral distributions.

\begin{figure*}[htp]
    \centering
    \includegraphics[width=0.8\linewidth]{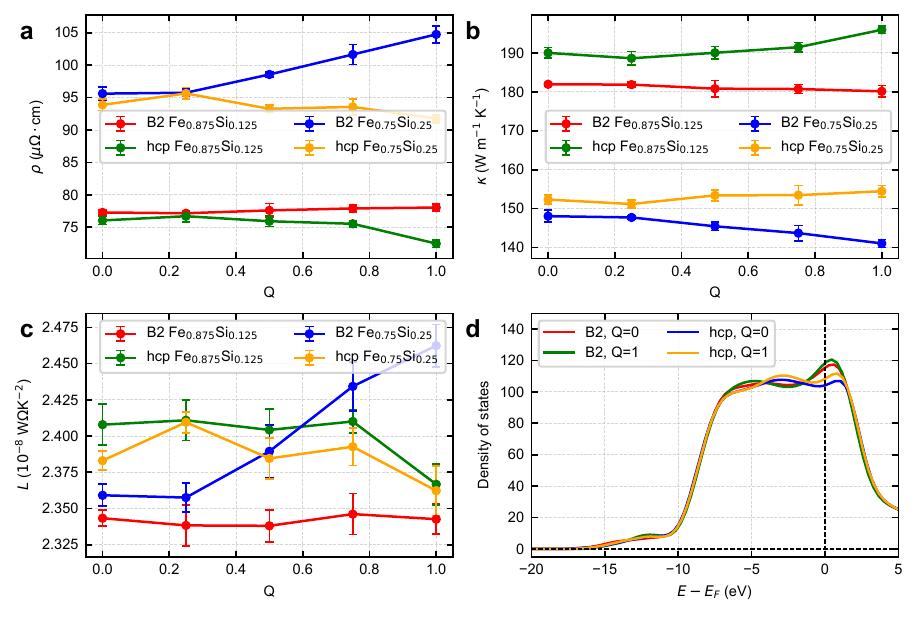}
    \caption{\textbf{Electrical transport properties of Fe–Si alloys at $\sim$330~GPa and 6000~K.} 
    \textbf{a} Electrical resistivity ($\rho$), \textbf{b} thermal conductivity ($\kappa$), and \textbf{c} Lorenz number ($L$) for B2 and hcp Fe–Si alloys with Si contents of 12.5~mol\% and 25~mol\% (Fe$_{0.875}$Si$_{0.125}$ and Fe$_{0.75}$Si$_{0.25}$). Error bars indicate the standard deviation among these four snapshots. The order parameter $Q$ varies from 0 (bcc, disordered) to 1 (B2, ordered). \textbf{d} Total electronic density of states (DOS) of Fe$_{0.875}$Si$_{0.125}$, with the Fermi energy shifted to zero. In the B2 structure, Fe$_{0.875}$Si$_{0.125}$ stabilizes in the disordered ($Q=0$, bcc-like) phase, while Fe$_{0.75}$Si$_{0.25}$ remains ordered ($Q=1$) at 6000 K.}
    \label{fig:transport}
\end{figure*}

\subsection*{Transport properties}

We calculated the electronic transport properties of Fe$_{0.875}$Si$_{0.125}$
alloy in both B2 and hcp structures at 330~GPa and 6000~K, corresponding
to Earth's inner-core (IC) conditions (Fig.~\ref{fig:transport}). Previous
experiments \cite{zhangThermalConductivityFeSi2022} demonstrated that Si
doping substantially increases the electrical resistivity of pure iron through
enhanced impurity scattering. 

Similarly, chemical ordering and structural changes in Fe--Si alloys modify
electronic scattering and consequently affect transport properties. Although
the total electronic density of states near the Fermi level changes only
slightly with the order parameter $Q$ (Fig.~\ref{fig:transport}d), the
electrical and thermal transport coefficients show pronounced sensitivity to
atomic ordering and crystal structure.

At 6000~K, the ordered B2 structure exhibits a slightly higher resistivity
($\sim$5.0~$\mu\Omega\cdot$cm) and lower thermal conductivity
($\sim$15~W~m$^{-1}$~K$^{-1}$) than the hcp structure.

Here and throughout the manuscript, B2 refers specifically to the
long-range-ordered state ($Q=1$), whereas the disordered cubic state is
denoted as bcc ($Q=0$).

The resulting Lorenz number of $\sim$2.3--2.5$\times$10$^{-8}$~W$\Omega$~K$^{-2}$
agrees well with the ideal Lorenz number
($L_0=2.45\times10^{-8}$~W$\Omega$K$^{-2}$)
(Fig.~\ref{fig:transport}c). For comparison, previous work
\cite{liuElectricalResistivityThermal2025} estimated a thermal conductivity
of $\sim$190~W~m$^{-1}$~K$^{-1}$ for FeH$_x$ alloys under IC conditions,
similar to that of the ordered hcp Fe$_{0.875}$Si$_{0.125}$ alloy
($\sim$197~W~m$^{-1}$~K$^{-1}$). In contrast, the disordered bcc
Fe$_{0.875}$Si$_{0.125}$ ($Q=0$) alloy has a lower thermal conductivity of
$\sim$182~W~m$^{-1}$~K$^{-1}$, about 7\% less than that of the corresponding
hcp phase.

We also conducted transport property calculations for the higher-silicon
composition Fe$_{0.75}$Si$_{0.25}$, with results shown in
Fig.~\ref{fig:transport}. Increasing the silicon content from
Fe$_{0.875}$Si$_{0.125}$ to Fe$_{0.75}$Si$_{0.25}$ induces pronounced changes
in electronic transport behavior. The electrical resistivity increases
substantially, while the thermal conductivity decreases by more than
$\sim$40~W~m$^{-1}$~K$^{-1}$, representing a reduction of about 30\%
compared with the lower-silicon alloy. Notably, the relationship between
electrical resistivity and the order parameter $Q$ differs between structures:
in the B2 phase, resistivity decreases with increasing $Q$ (a positive
correlation), whereas in the hcp phase, it shows a weak negative correlation.
This contrasting behavior highlights the interplay between chemical disorder
and crystal symmetry in the transport properties of Fe--Si alloys under
inner-core conditions.

Electrical and thermal conductivities of iron alloys under core conditions
provide important constraints for models of Earth's thermal evolution.
However, translating the calculated transport contrasts into changes in core
cooling rates, inner-core growth, or geodynamo evolution requires additional
information, including the abundance, spatial distribution, and lifetime of
different phases, together with a coupled thermal-evolution model.

Our calculations demonstrate that order--disorder transitions and structural
changes produce measurable variations in electronic transport. In particular,
the reduced thermal conductivity of disordered bcc Fe$_{0.875}$Si$_{0.125}$
suggests that possible metastable or localized cubic components could modify
heat transport relative to an entirely hcp inner core. At higher Si
concentrations, stabilization of ordered B2 further changes the transport
response. These results provide phase-dependent transport constraints for
future inner-core evolution models rather than a direct prediction of core
cooling history or magnetic-field evolution.

\begin{figure*}
    \centering
    \includegraphics[width=0.8\linewidth]{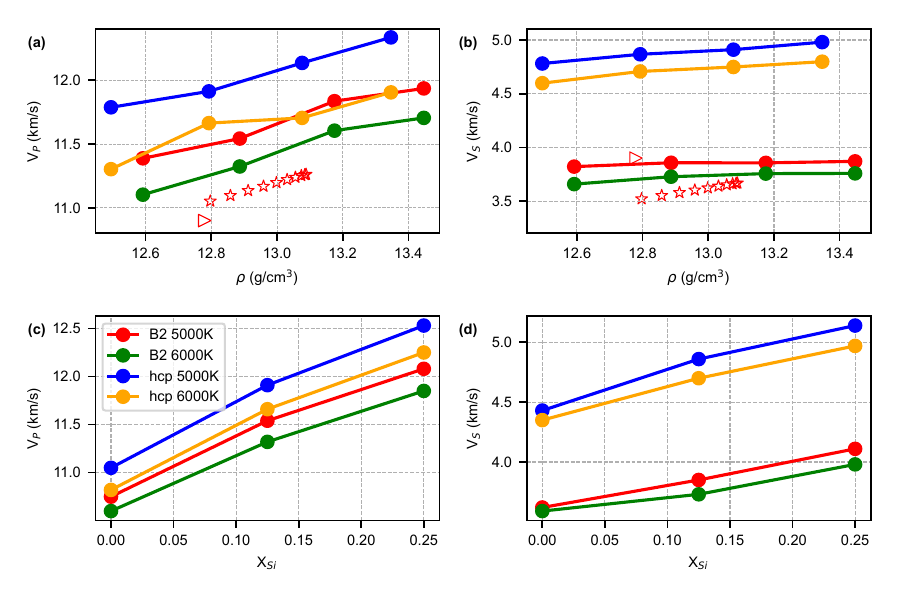}
    \caption{\textbf{Seismic velocities of Fe–Si alloys as a function of density and composition.} Calculated compressional-wave ($V_P$) and shear-wave ($V_S$) velocities of Fe$_{0.875}$Si$_{0.125}$ alloys in the B2 (\textbf{a}) and hcp (\textbf{b}) phases are compared with the Preliminary Reference Earth Model (PREM) \cite{dziewonskiPreliminaryReferenceEarth1981} values for the inner core (open red stars) and with previous simulations (open triangles)\cite{liShortrangeOrderStabilizes2025}. Panels (\textbf{c}) and (\textbf{d}) show the composition dependence of seismic velocities for B2 and hcp phases at 5000~K and 6000~K, including pure Fe, Fe$_{0.875}$Si$_{0.125}$, and Fe$_{0.75}$Si$_{0.25}$. In the B2 structure, Fe$_{0.875}$Si$_{0.125}$ is stabilized in the disordered ($Q=0$, bcc-like) phase, whereas Fe$_{0.75}$Si$_{0.25}$ remains ordered ($Q=1$) at 5000~K and 6000~K.The average pressures from the FPMD trajectories at 6000~K are 338.2~GPa for disordered bcc Fe$_{0.875}$Si$_{0.125}$ and 343.2~GPa for hcp Fe$_{0.875}$Si$_{0.125}$; the full pressure dependence is provided in Fig.~S10.}
    \label{fig:seismic}
\end{figure*}

\subsection*{Seismic properties}
In addition to electrical transport properties, the crystal structure (B2 vs. hcp) also has a strong influence on the elastic constants and seismic velocities of Fe–Si alloys under inner core (IC) conditions. The calculated elastic constants ($C_{ij}$) decrease systematically with increasing temperature. Although both structures exhibit comparable bulk moduli, the disordered bcc ($Q=0$) Fe$_{0.875}$Si$_{0.125}$ alloy possesses a substantially lower shear modulus than the hcp phase, resulting in significantly reduced shear wave velocities under IC conditions (Table~S2 in the Supplemental Materials). 

The origin of seismic anisotropy in Earth’s inner core remains debated, with recent studies proposing diverse mechanisms to explain the observed seismic signatures. The interplay between seismological observations and mineral physics models suggests that inner-core anisotropy may arise from depth-dependent phase transitions, shear softening, and variations in light-element composition \cite{wangSupportEquatorialAnisotropy2018,frostOrientationFastSlow2019,wangShearWaveAnisotropyEarths2021,tkalcicShearPropertiesEarths2022,kolesnikovStrengthSeismicAnisotropy2022,wuShearSofteningEarths2025}. Consistent with previous results for pure hcp Fe \cite{shaElasticIsotropyEFe2010}, our calculations show that the elastic anisotropy of hcp Fe–Si alloys is nearly zero under inner-core conditions, indicating that the hcp phase alone cannot account for the observed seismic anisotropy of Earth’s inner core.  

We computed the directional dependence of elastic anisotropy for Fe$_{0.875}$Si$_{0.125}$ at 5000~K and 6000~K (Fig.~S9 in the Supplemental Materials). For both compressional and shear waves propagating along different orientations relative to the $c$-axis, the B2 structure exhibits stronger velocity variations than the hcp phase. At 6000~K, the shear wave velocity in B2 Fe–Si ranges from $\sim$2.4 to 5.0~km/s, corresponding to a pronounced Chung–Buessem shear anisotropy of 22.9\%, compared with only 0.6\% for the hcp phase. These values represent ideal single-crystal limits; in the real inner core, polycrystalline averaging due to solidification and plastic deformation would substantially reduce the observable anisotropy. Nevertheless, the intrinsically higher anisotropy of the B2 structure suggests that even partial crystallographic alignment during inner-core growth could produce the seismic anisotropy detected in observations. Together, these results suggest that B2 Fe–Si alloys—stabilized by elevated temperature and silicon enrichment—may play a crucial role in reconciling mineral physics predictions with the observed seismic anisotropy of Earth's inner core.

The compressional ($V_P$) and shear ($V_S$) wave velocities of disordered B2 and hcp Fe$_{0.875}$Si$_{0.125}$ alloys were calculated at varying densities using their elastic constants, as shown in Fig.~\ref{fig:seismic} \textbf{a},\textbf{b}. Both $V_P$ and $V_S$ increase with density, consistent with the effects of compression. Previous studies \cite{liShortrangeOrderStabilizes2025} reported a $V_S$ of 3.93~km/s at 12.78~g/cm$^3$ for a short-range ordered stabilized bcc Fe–Si alloy, slightly higher than our calculated value of 3.73~km/s. Notably, the $V_S$ curve of the B2 phase at 6000~K aligns more closely with the PREM model than that of the hcp phase, suggesting that the B2 structure—potentially stabilized by silicon incorporation—is a compelling candidate for the dominant Fe-bearing phase in Earth’s inner core. The pressure dependence of seismic velocities is shown in Fig.~S10 of the Supplemental Materials.

Beyond Fe$_{0.875}$Si$_{0.125}$, we extended our investigation to pure Fe and Fe$_{0.75}$Si$_{0.25}$ compositions at 5000~K and 6000~K (Fig.~\ref{fig:seismic} \textbf{c},\textbf{d}). Both $V_P$ and $V_S$ increase approximately linearly with silicon content, consistent with previous experimental and theoretical studies \cite{hiroseLightElementsEarths2021a}. By interpolating to the geophysically plausible Si range of 4.5–7.9~mol\% \cite{liuHydrogenSiliconAre2024,hiroseLightElementsEarths2021a}, we estimate shear-wave velocities at the inner-core boundary (ICB) of 3.68–3.70~km/s for the B2 phase and 4.64–4.67~km/s for the hcp phase. The hcp values are notably higher than the PREM shear velocity of $\sim$3.6~km/s, reinforcing the long-recognized discrepancy between seismological observations and models based solely on hcp phase.

We further derived Poisson’s ratio from the calculated $V_P$ and $V_S$ values (Fig.~S11 in the Supplemental Materials). The density- and composition-dependent Poisson’s ratio of the bcc-like Fe$_{0.875}$Si$_{0.125}$ alloy is significantly higher than that of the hcp phase and closely matches seismological estimates ($\nu = 0.44$)\cite{dziewonskiPreliminaryReferenceEarth1981}. In particular, the B2 phase exhibits a $V_P/V_S$ ratio of $\sim$3.0 and a Poisson’s ratio of $\sim$0.44, about 20\% and 10\% greater than those of the hcp phase, respectively. Previous studies\cite{hiroseLightElementsEarths2021a,liElasticPropertiesHcpFe2018a} constrained the silicon content of the inner core to relatively low values, assuming that Si is incorporated only into hcp Fe alloys. However, the hcp Fe–Si phase alone cannot reproduce the observed high $V_P/V_S$ ratio. Our results therefore suggest that incorporating the B2 Fe–Si phase yields somewhat higher Si concentrations in the inner core, while remaining consistent with seismological constraints.

Overall, our calculations show that cubic Fe--Si exhibits substantially lower
shear-wave velocities and higher $V_P/V_S$ ratios than hcp Fe--Si under
inner-core conditions. In particular, the disordered bcc state at low Si
content yields seismic velocities closer to PREM than the corresponding hcp
phase. Therefore, the presence of a cubic component within an otherwise
hcp-dominated inner core could help reduce the discrepancy between hcp-based
mineral-physics predictions and observed seismic velocities. The magnitude of
this contribution will depend on the cubic phase fraction, composition, and
crystallographic texture.

Seismological studies have also suggested radial variations in inner-core
anisotropy, including possible differences between the innermost and overlying
inner core
\cite{wilsonFormationEvolutionEarths2025,wangEquatorialAnisotropyInner2015}.
As discussed above, the density contrast between coexisting hcp and cubic
Fe--Si (Fig.~S14 in the Supplemental Materials) provides a possible
mechanism for depth-dependent variations in their relative abundances. Because disordered bcc Fe--Si exhibits
substantially larger intrinsic single-crystal shear anisotropy than hcp
Fe--Si, such a radial variation in phase fraction could contribute to
depth-dependent seismic anisotropy.

In summary, we constructed a two-site model to describe the evolution from
ordered B2 to disordered bcc states across varying Si concentrations in
Fe--Si alloys. First-principles calculations reveal composition- and
temperature-dependent order--disorder transitions arising from the interplay
of enthalpy, configurational entropy, and vibrational entropy. Using the
resulting free energies, we constructed the Fe--Si binary phase diagram from
3000 to 6000~K and identified a broad hcp--cubic two-phase field under
inner-core conditions.

At geophysically plausible bulk Si concentrations, hcp remains the predominant
equilibrium phase, whereas cubic Fe--Si may occur as a
composition-dependent, metastable, or spatially variable component. Relative
to hcp Fe--Si, the cubic phase exhibits lower shear-wave velocities, larger
intrinsic elastic anisotropy, and distinct electronic transport properties.
Together with the density contrast between the coexisting phases, these
results provide thermodynamic, seismic, and transport constraints on the
possible coexistence and spatial distribution of hcp and cubic Fe--Si phases
in Earth's inner core.

\bibliography{main.bib}

\section*{Acknowledgments}
We thank Russell J. Hemley for insightful discussions and valuable suggestions that greatly improved this work. This work is supported by the US National Science Foundation CSEDI grant EAR-1901813 and the Carnegie Institution for Science. We gratefully acknowledge supercomputer support from the Resnick High Performance Computing Center. The authors gratefully acknowledge the Gauss Centre for Supercomputing e.V. (\href{http://www.gauss-centre.eu/}{www.gauss-centre.eu}) for funding this project by providing computing time on the GCS Supercomputer SuperMUC-NG at Leibniz Supercomputing Centre (\href{http://www.lrz.de/}{www.lrz.de}).
\\
\section*{Additional information}
Supplemental Information is available in the online version of the paper.
\\
\section*{Competing financial interests}
The authors declare no competing financial interests.
\\
\section*{Author Contributions Statement}
C.L. designed the research and performed the simulations. C.L., X.D., and R.E.C. contributed to data processing, numerical simulations, and analysis. C.L. wrote the first draft of the manuscript, which X.D. and R.E.C. subsequently reviewed and improved.
\\
\\
\section*{Code Availability}
VASP is a commercial software package available from https://www.vasp.at. Detailed information on access, installation, and usage of these codes are provided on their respective websites. The input files used in this study will be publicly available on Zenodo.

\onecolumngrid
\newpage

\setcounter{figure}{0}

\renewcommand \thesection{S\@arabic\c@section}
\renewcommand\thetable{S\@arabic\c@table}
\renewcommand \thefigure{S\@arabic\c@figure}
\makeatother

\renewcommand{\thefigure}{S\arabic{figure}}
\renewcommand{\figurename}{Fig.}
\renewcommand{\thetable}{S\arabic{table}}
\renewcommand{\tablename}{Table.}

\begin{center}
    {\Large \textit{Supplemental Materials}}\\[1.2ex]
    {\large \bf Order–Disorder in Fe–Si Alloys: Implications for Seismic and Transport Properties of Earth's Inner Core}\\[2.0ex]
    {\normalsize
    Cong Liu$^{1,\ast}$, Xin Deng$^{1}$, and R.~E. Cohen$^{1,\dagger}$\\[1.2ex]
    $^{1}$Extreme Materials Initiative, Earth and Planets Laboratory, Carnegie Institution for Science,\\
    5241 Broad Branch Road NW, Washington, DC 20015, USA\\[0.8ex]
    
    $^\ast$Corresponding author: cliu10@carnegiescience.edu\\
    $^\dagger$rcohen@carnegiescience.edu
    }
\end{center}

\section*{Methods}
\subsection*{Configuration generation}

To investigate the order–disorder phase transition in Fe–Si alloys, we introduced an order parameter $Q$ within a two-site model framework for bcc, hcp, and fcc structures (Fig.~S1 in the Supplemental Materials). 
The order parameter quantifies the degree of chemical ordering and is defined as follows:
\begin{equation}
    Q = 
    \frac{\sum_i \left| C_i^{\mathrm{struct}} - C_i^{\mathrm{random}} \right|}
         {\sum_i \left| C_i^{\mathrm{random}} \right|},
\end{equation}
where $C_i^{\mathrm{struct}}$ and $C_i^{\mathrm{random}}$ represent the correlation functions for a given structure and for a perfectly random alloy, respectively. Here, $Q=0$ corresponds to a fully disordered configuration with Si atoms equally distributed between both sites, while $Q=1$ denotes a fully ordered phase with Si atoms occupying only one sublattice.

We generated $Q$-dependent atomic configurations using the Alloy Theoretic Automated Toolkit (ATAT) with the Special Quasirandom Structure (SQS) method\cite{vandewalleAlloyTheoreticAutomated2002}. 
An SQS is a finite periodic supercell that approximates the correlation functions (pairs, triplets, etc.) of a perfectly random alloy within a limited number of atoms.
Pair clusters up to the second, third, and fourth nearest neighbors (cutoff distances 2.5, 3.0, and 3.5~\AA) were included, and the deviation from random correlation functions was minimized. The SQS thus provides a reproducible and efficient representation of chemical disorder, enabling accurate evaluation of thermodynamic and electronic properties while capturing local lattice distortions.  SQS is especially appropriate at the high temperatures of planetary interiors since it represents disordered states, with possible long-range order. The SQS configurations used in this work represent controlled disordered
reference states at fixed order parameter $Q$. They reproduce selected pair
and multisite correlation functions of an ideal random alloy within a finite
supercell. Short-range order is not included, as is appropriate at the high temperatures of planetary cores.

For each target order parameter $Q$, we constructed a 128-atom ($4\times4\times4$) supercell for the bcc, hcp, and fcc lattices. The procedure included: (i) defining the parent lattice and atomic basis in a VASP-compatible \texttt{POSCAR}; (ii) specifying the target composition; (iii) performing Monte Carlo sampling to generate candidate configurations; (iv) evaluating the deviation of correlation functions from random values; and (v) selecting the configuration with the lowest score as the representative SQS. The SQS quality was validated by comparing short-range pair and multi-site correlations to those of a random alloy.

We systematically varied $Q$ from 0 to 1 for each structure type and Si concentration, using 5, 5, 4, and 5 intermediate $Q$ values for Si = 12.5, 25.0, 37.5, and 50.0~mol\%, respectively. Atomic occupation details for each configuration are listed in Table~S1 of the Supplemental Materials.

The configurational entropy $S_\mathrm{conf}$ was calculated as
\begin{equation}
    \frac{S_\mathrm{conf}}{k_B} = -\sum X \ln X,
\end{equation}
where $X$ is the site occupation probability of Fe or Si and $k_B$ is Boltzmann’s constant. At a fixed composition, $S_\mathrm{conf}$ decreases monotonically as $Q$ increases from 0 to 1. The entropy difference between ordered and disordered states depends strongly on Si concentration (Fig.~S5 in the Supplemental Materials).

Convergence tests using larger supercells (up to 200 atoms) confirmed that both
the free energy and order parameter $Q$ converge within statistical
uncertainty, indicating that a 128-atom supercell provides a reliable
representation of the disordered phase (Fig.~S12 in the Supplemental
Materials). Specifically, for Fe$_{0.5}$Si$_{0.5}$ at 330~GPa, the enthalpy
decreases from the disordered bcc phase ($Q=0$) toward the ordered B2 phase
($Q>0$), with larger supercells improving convergence, particularly at lower
$Q$. Molecular dynamics simulations of the disordered bcc structure at 6000~K
show that the Gibbs free energy plateaus beyond 108 atoms, with variations
within 10~meV/atom between 128 and 200 atoms. Thus, the 128-atom supercell
offers a robust compromise between accuracy and computational efficiency.

To further evaluate the sensitivity to the choice of SQS realization, five
independent 128-atom SQS configurations were examined for disordered bcc and
hcp Fe$_{0.875}$Si$_{0.125}$ at 330~GPa. The resulting static enthalpy
variations are within 10~meV/atom, indicating limited configurational
uncertainty associated with the selected SQS representations. This analysis
constrains the variability of the static energy among representative SQS
configurations, but it does not constitute a complete sampling of vibrational
entropy or finite-temperature configurational fluctuations.

\subsection*{First-principles calculations}

The optimal SQS structure, corresponding to the minimum value of the objective function, was optimized at 330~GPa through full relaxation of both lattice parameters and atomic positions. For the bcc and fcc structures, the cell shape was constrained by fixing the lattice constants such that $a = b = c$, whereas for the hcp structure, the $c/a$ ratio was allowed to relax.

Density functional theory (DFT) calculations were performed using the Vienna \textit{Ab initio} Simulation Package (VASP)~\cite{kresseEfficientIterativeSchemes1996}. The projector augmented-wave (PAW) method~\cite{blochlProjectorAugmentedwaveMethod1994} was employed together with the generalized gradient approximation (GGA) using the Perdew–Burke–Ernzerhof (PBE) exchange–correlation functional~\cite{perdewGeneralizedGradientApproximation1996a}. The $3s^2 3p^6 4s^2 3d^6$ electrons in Fe and the $3s^2 3p^2$ electrons in Si were treated as valence electrons. A plane-wave energy cutoff of 500~eV was adopted, and Brillouin-zone integrations were carried out using a $2\times2\times2$ Monkhorst–Pack $k$-point mesh. The total energy convergence threshold for structural relaxation was set to $10^{-6}$~eV.

\subsection*{First-principles molecular dynamics simulations}

To account for the vibrational entropy of Fe–Si alloys (128 atoms) under high-pressure and high-temperature conditions, first-principles molecular dynamics (FPMD) simulations were performed based on the Born–Oppenheimer approximation, as implemented in VASP. The Perdew–Burke–Ernzerhof (PBE) exchange–correlation functional within the generalized gradient approximation was employed, with Brillouin-zone sampling restricted to the $\Gamma$ point. A plane-wave energy cutoff of 500~eV was used, ensuring total-energy convergence better than $10^{-5}$~eV.

Equilibrium volumes at different temperatures were determined by conducting NPT simulations over a grid of temperatures at 330 GPa, using a Langevin thermostat~\cite{allenComputerSimulationLiquids2017a} to control temperature. Each molecular dynamics simulation was run for 6~ps with a time step of 1.0~fs, of which the first 1~ps was discarded for equilibration. Data from the remaining 5~ps were used to compute ensemble-averaged properties. The internal energies of Fe–Si alloys at various temperatures were obtained from these averages, and the associated uncertainties were propagated through the construction of common tangents on the Gibbs free energy curves, thereby constraining the phase boundaries (Fig.3).

\subsection*{Chemical potential calculations}
To represent the composition- and temperature-dependent Gibbs free energy of the Fe--Si system, we adopted a combined polynomial--logarithmic expression:
\begin{equation}
\begin{aligned}
&G(X, T) \\
=&\;  a + bX + cX^2 + dX^3 + eX^4 \\
            & + k_BT\left[\, X\ln X + (0.5 - X)\ln(0.5 - X) \,\right]
\label{eq:G_model}
\end{aligned}
\end{equation}

In this formulation, \(X\) denotes the atomic fraction of Si (limited to \(X \leq 0.5\)). The polynomial terms capture the enthalpic contributions to the free energy, while the logarithmic term describes the ideal configurational entropy of mixing in a binary Fe--Si solution. This model accurately reproduces our computed Gibbs free energy data, with fitting residuals below 1~meV/atom across the sampled temperature and composition range.

The chemical potential difference between Si and Fe is given by:
\begin{equation}
\Delta\mu = \mu_{\mathrm{Si}} - \mu_{\mathrm{Fe}} = \frac{\partial G(X, T)}{\partial X}.
\label{eq:mu_diff_def}
\end{equation}

Differentiating Eq.~\ref{eq:G_model} with respect to \( x \) yields:
\begin{equation}
\Delta\mu = b + 2cX + 3dX^2 + 4eX^3 + k_BT \ln\left( \frac{X}{0.5 - X} \right).
\label{eq:mu_diff_final}
\end{equation}

\subsection*{Vibrational density of states and vibrational entropy}

The vibrational density of states (VDOS) was obtained from the Fourier transform of the velocity autocorrelation function (VACF) computed from the FPMD trajectories. The VACF is defined as  
\begin{equation}
    C(t) = \frac{1}{N} \sum_{i=1}^{N} \langle \mathbf{v}_i(0) \cdot \mathbf{v}_i(t) \rangle,
\end{equation}
where \( \mathbf{v}_i(t) \) is the velocity of atom \( i \) at time \( t \), and the average is taken over all \( N \) atoms and multiple time origins along the trajectory. The VDOS \( g(\omega) \) is then obtained as the Fourier transform of the normalized VACF:
\begin{equation}
    g(\omega) = \int_0^{\infty} C(t) \, e^{-i\omega t} \, dt.
\end{equation}
In practice, the VACF is computed from discretely sampled velocities, and the Fourier transform is evaluated using a fast Fourier transform (FFT). The resulting VDOS is normalized according to
\begin{equation}
    \int_0^{\infty} g(\omega) \, d\omega = 3.
\end{equation}

This approach inherently captures anharmonic vibrational effects at finite temperatures, providing a realistic description of lattice dynamics under extreme thermodynamic conditions relevant to the Earth’s inner core. The Fe–Si alloys exhibit solid-like atomic motion in all simulations, as indicated by non-diffusive mean-square displacements (MSD; Fig.~S13 in the Supplemental Materials). Furthermore, the vanishing low-frequency limit of the VDOS, $g(0) = 0$, confirms the absence of diffusive behavior and the preservation of solid phases.

The vibrational entropy per atom at temperature \( T \) is computed under the quasi-harmonic approximation as
\begin{equation}
\begin{aligned}
    S_{\mathrm{vib}} = 
    k_B \int_0^{\infty} g(\omega) \Bigg[
        &\frac{\hbar \omega}{k_B T}
        \cdot \frac{1}{e^{\hbar \omega / k_B T} - 1} \\
        &- \ln\!\left(1 - e^{-\hbar \omega / k_B T}\right)
    \Bigg] d\omega,
\end{aligned}
\end{equation}
where \( k_B \) is the Boltzmann constant and \( \hbar \) is the reduced Planck constant. This expression accounts for phonon occupation following Bose–Einstein statistics and incorporates the temperature dependence of vibrational entropy.

The zero-point energy (ZPE), arising from quantum zero-point motion, corresponds to the vibrational energy at absolute zero temperature (\( T = 0 \)~K) and is calculated as
\begin{equation}
    E_{\mathrm{ZPE}} = \int_0^{\infty} \frac{1}{2} \hbar \omega \, g(\omega) \, d\omega.
\end{equation}
This quantity represents the residual vibrational energy retained by the atoms due to quantum effects even at 0~K.

\subsection*{Transport property calculations}

The electronic transport properties of the CH compound were computed using a modified spin-polarized relativistic Korringa–Kohn–Rostoker (SPR-KKR) code~\cite{ebertCalculatingCondensedMatter2011,minarCorrelationEffectsTransition2011}. Four uncorrelated atomic snapshots were randomly selected from \textit{ab initio} molecular dynamics (AIMD) simulations performed under high-pressure and high-temperature conditions.

For each snapshot, a self-consistent field (SCF) calculation was carried out within the framework of density functional theory (DFT), employing a maximum angular momentum cutoff of $l_{\mathrm{max}} = 3$ and 125 $k$-points in the Brillouin zone. After convergence, the transport properties were evaluated using the Kubo–Greenwood formalism with the same $l_{\mathrm{max}}$, a denser $k$-point mesh of 216 points, and an energy window of $[-5k_{\mathrm{B}}T, +5k_{\mathrm{B}}T]$ centered around the chemical potential.

The electrical conductivity ($\sigma$), resistivity ($\rho = 1/\sigma$), and thermal conductivity ($\kappa$) were derived from the transport coefficients $\mathcal{L}_{ij}$ according to
\begin{equation}
    \sigma = \mathcal{L}_{11}, \quad
    \kappa = \frac{1}{eT}
    \left(
        \mathcal{L}_{22} - \frac{\mathcal{L}_{12}^2}{\mathcal{L}_{11}}
    \right),
\end{equation}
where the transport coefficients are defined as
\begin{equation}
    \mathcal{L}_{ij} =
    (-1)^{i+j} \int d\varepsilon \,
    \sigma_{\mu\nu}(\varepsilon - \mu)^{i + j - 2}
    \left( -\frac{\partial f_T(\varepsilon)}{\partial \varepsilon} \right).
\end{equation}
Here, $f_T(\varepsilon)$ is the Fermi–Dirac distribution at temperature $T$, and $\mu$ is the chemical potential.

The energy-dependent conductivity tensor $\sigma_{\mu\nu}(\varepsilon)$ was evaluated from the Kubo–Greenwood expression,
\begin{equation}
    \sigma_{\mu \nu} \propto
    \operatorname{Tr}
    \left\langle
        \hat{j}_\mu \,
        \mathrm{Im}G^+ \,
        \hat{j}_\nu \,
        \mathrm{Im}G^+
    \right\rangle,
\end{equation}
where $\hat{j}_\mu$ is the current operator and $\mathrm{Im}G^+$ denotes the imaginary part of the retarded Green’s function.

\subsection*{Elastic property calculations}

The independent elastic constants $C_{ij}$ ($i,j = 1, 2, \ldots, 6$) were determined by applying small finite distortions to the equilibrium structure and solving the linear stress–strain relations. These constants provide the foundation for evaluating the seismic velocities and elastic anisotropy of Fe–Si alloys under Earth’s inner-core conditions.

The elastic properties of disordered bcc- and hcp-structured Fe$_{0.875}$Si$_{0.125}$ alloys were computed under high-pressure and high-temperature conditions using first-principles molecular dynamics (FPMD) simulations. For each structure and temperature, the simulations were run for at least 6~ps to achieve equilibrium, after which the stress tensor was averaged over the final 5~ps of the trajectory.

To evaluate the elastic constants, a set of finite strains $\varepsilon$ with magnitudes $\delta = 0$, $\pm0.005$, and $\pm0.01$ was applied to the relaxed structures, generated using the VASPKIT package~\cite{WANG2021108033}. The corresponding stress tensors $\sigma_{ij}$ were obtained from DFT calculations for each strained configuration. The elastic constants $C_{ij}$ were then derived by linearly fitting the stress–strain data according to the generalized Hooke’s law:
\begin{equation}
    \sigma_{ij} = \sum_{k,l} C_{ijkl} \, \varepsilon_{kl}.
\end{equation}

After obtaining $C_{ij}$, the bulk modulus $K$ and shear modulus $G$ were evaluated using the Voigt–Reuss–Hill averaging scheme. The compressional ($V_P$) and shear ($V_S$) wave velocities were then calculated as
\begin{align}
    V_P &= \sqrt{\frac{K + \tfrac{4}{3}G}{\rho}}, \\
    V_S &= \sqrt{\frac{G}{\rho}},
\end{align}
where $\rho$ is the density at the target pressure.

The Chung–Buessem shear anisotropy factor $A_{\mathrm{CB}}$ is defined as
\begin{equation}
    A_{\mathrm{CB}} = \frac{G_V - G_R}{G_V + G_R},
\end{equation}
where $G_V$ and $G_R$ are the Voigt and Reuss averages of the shear modulus, respectively. $A_{\mathrm{CB}} = 0$ corresponds to perfect elastic isotropy, while larger values indicate increasing anisotropy. The elastic constants and derived quantities of Fe$_{0.875}$Si$_{0.125}$ at inner-core conditions are listed in Table~S2 in the Supplemental Materials.

The Poisson ratio $\nu$ was determined from the $V_P/V_S$ ratio as
\begin{equation}
    \nu = \frac{0.5\,(V_P/V_S)^2 - 1}{(V_P/V_S)^2 - 1}.
\end{equation}

\subsection*{Pure-Fe bcc--hcp free-energy benchmark}

At 330~GPa and 6000~K, we obtain a pure-Fe Gibbs free-energy difference of
approximately \(85~\mathrm{meV/atom}\) between bcc and hcp, compared with
approximately \(45~\mathrm{meV/atom}\) in the previous
thermodynamic-integration calculation
\cite{liShortrangeOrderStabilizes2025}. At these pressures,

\[
\Delta G_{\mathrm{bcc-hcp}}
=
\Delta F_{\mathrm{bcc-hcp}}
+
P\left(V_{\mathrm{bcc}}-V_{\mathrm{hcp}}\right).
\]

Our calculated equilibrium atomic volumes at 6000~K are
\(6.9017~\text{\AA}^{3}/\mathrm{atom}\) for bcc and
\(6.8739~\text{\AA}^{3}/\mathrm{atom}\) for hcp, giving

\[
P\Delta V_{\mathrm{bcc-hcp}}
\approx +57~\mathrm{meV/atom}.
\]

Together with
\(\Delta F_{\mathrm{bcc-hcp}}\approx28~\mathrm{meV/atom}\), this yields

\[
\Delta G_{\mathrm{bcc-hcp}}
\approx85~\mathrm{meV/atom}.
\]

The positive pressure--volume term therefore further disfavors bcc because
its equilibrium atomic volume is slightly larger than that of hcp. The
remaining difference relative to previous thermodynamic-integration results
is not uniquely resolved here. Importantly, the larger bcc--hcp separation in
our calculation disfavors rather than stabilizes the cubic phase; reducing
this separation toward the previous estimate would therefore favor bcc/B2.

\newpage
\section*{Supplemental figures and tables}
\begin{figure*}[htp]
    \centering
    \includegraphics[width=0.9\linewidth]{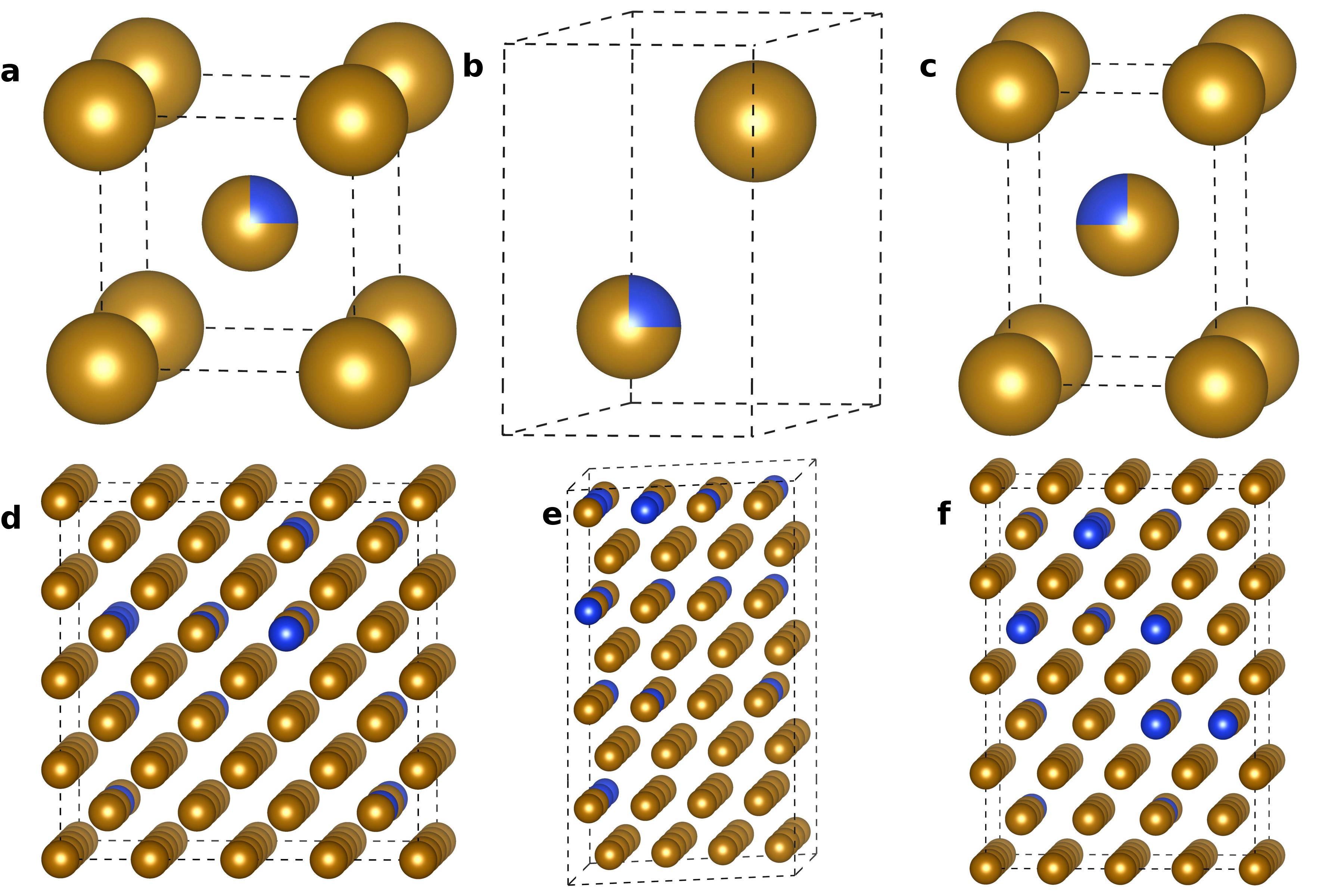}
    \caption{\textbf{Two-site model and corresponding SQS} for (a, d) B2, (b, e) hcp, and (c, f) fcc phases.
    The two-site model of the fcc phase is transformed from the four-atom conventional fcc cell. Iron and silicon atoms are shown in brown and blue, respectively. The figure shows fully ordered phases ($Q = 1$) at a composition of Fe$_{0.875}$Si$_{0.125}$, where Si atoms partially substitute on a specific crystallographic site.}
    \label{fig:atoms}
\end{figure*}

\begin{figure*}[htp]
    \centering
    \includegraphics[width=0.9\linewidth]{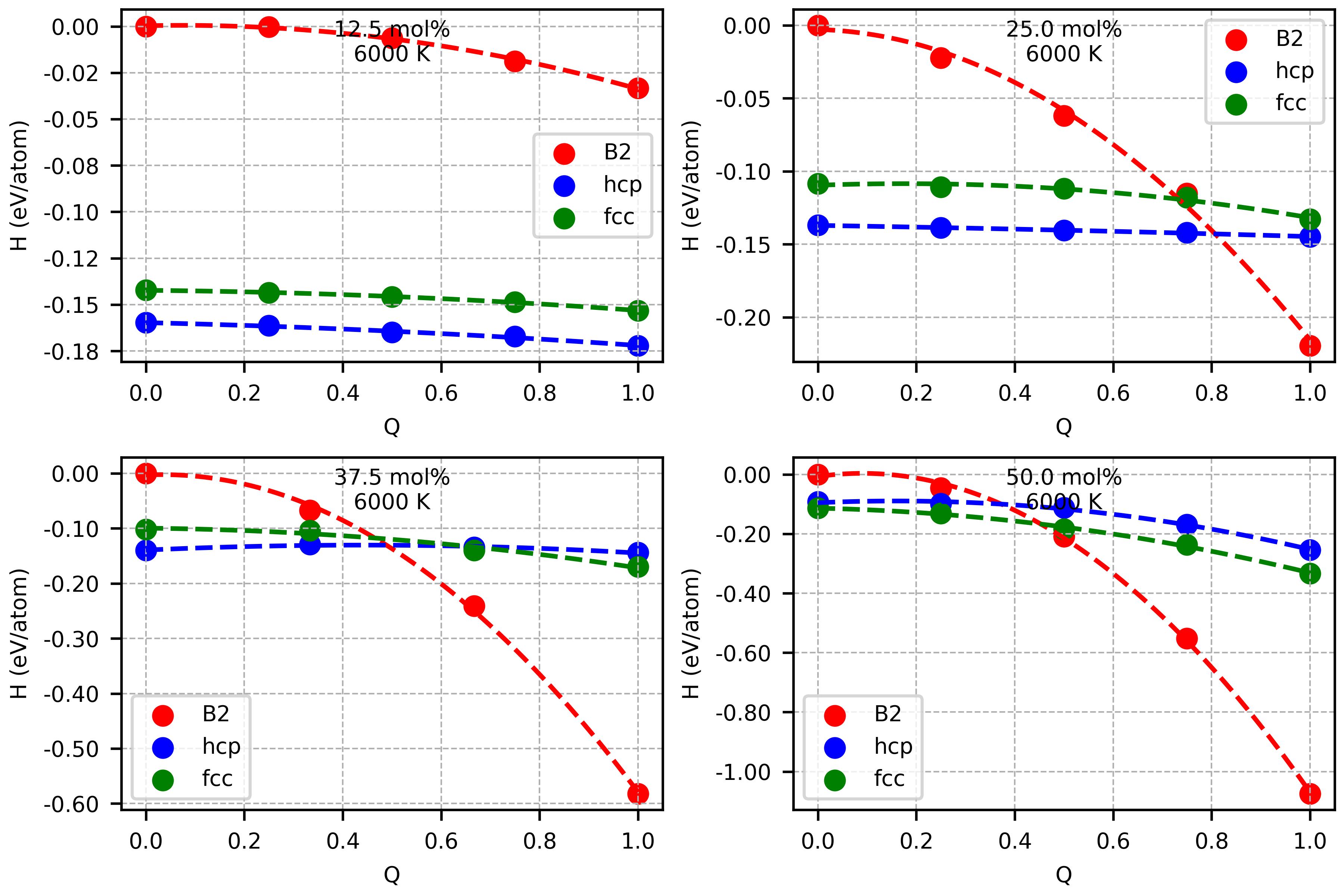}
    \caption{\textbf{Enthalpy of Fe–Si alloys at 330~GPa and 6000~K.}
    High-temperature enthalpies were obtained from molecular dynamics simulations, averaged over the final 5~ps of equilibrated trajectories. The order parameter $Q$ was varied for B2 (red), hcp (blue), and fcc (green) phases of Fe–Si alloys with Si content of 12.5, 25.0, 37.5, and 50.0~mol\%. Enthalpies of the B2 phase at $Q=0$ are shifted to zero for comparison.}
    \label{fig:H}
\end{figure*}

\begin{figure*}[htp]
    \centering
    \includegraphics[width=0.9\linewidth]{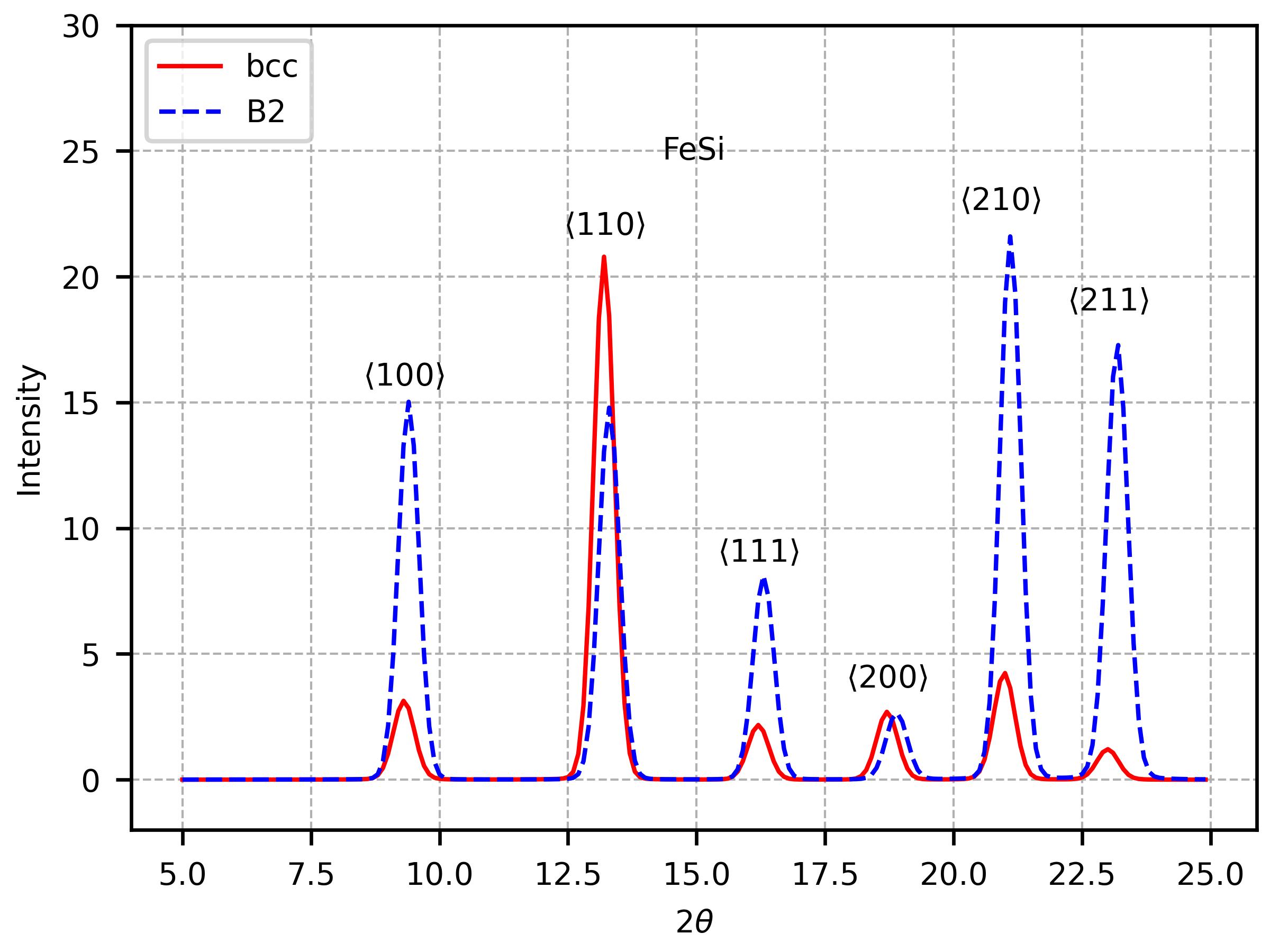}
    \caption{\textbf{Powder diffraction patterns of bcc/B2 phase Fe$_{0.875}$Si$_{0.125}$ alloys at Earth's inner-core conditions.}
    X-ray diffraction data were derived from molecular dynamics snapshots of disordered (solid red) and ordered (dashed blue) structures.}
    \label{fig:xrd}
\end{figure*}

\begin{figure*}[htp]
    \centering
    \includegraphics[width=0.9\linewidth]{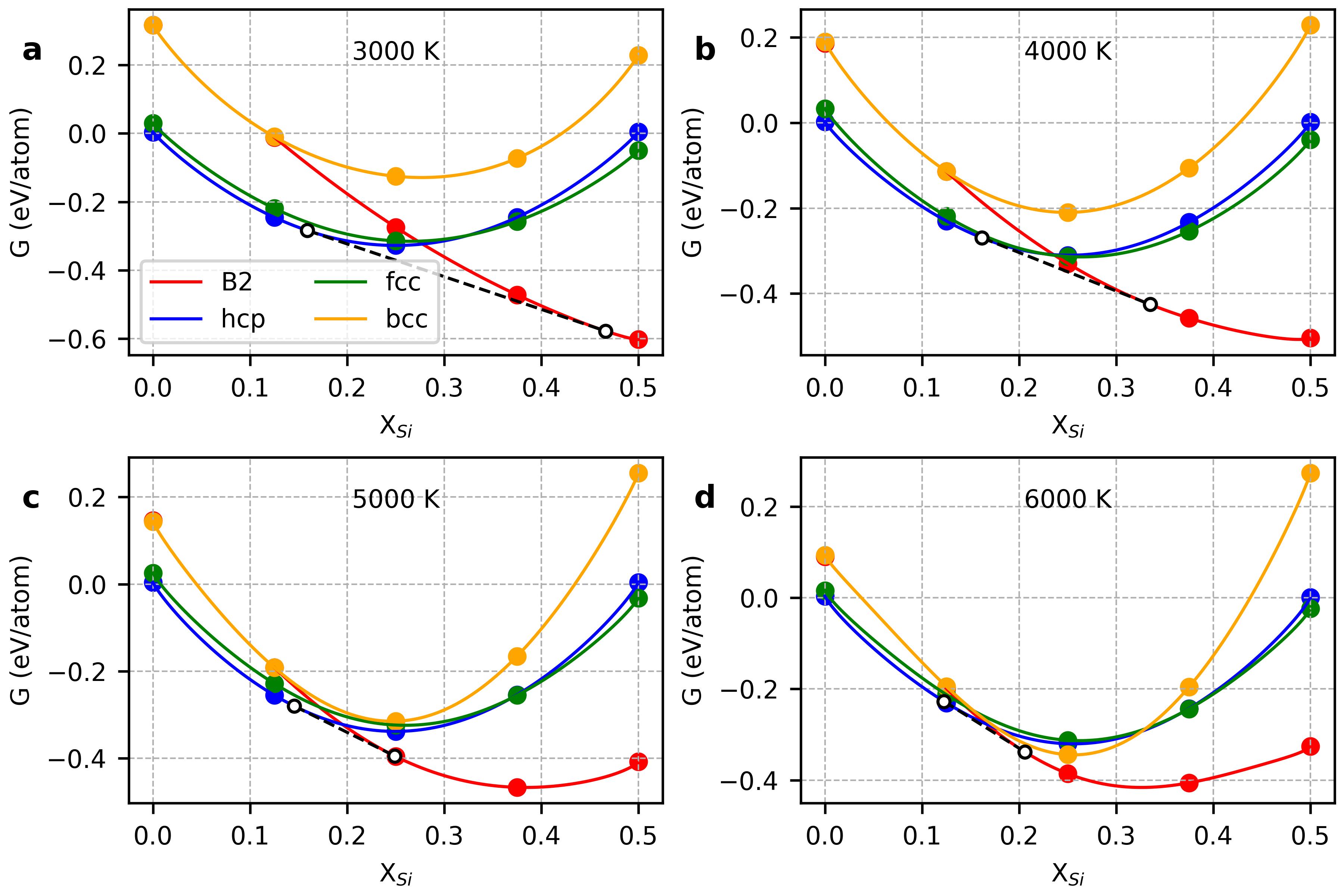}
    \caption{\textbf{Gibbs free energy $G$ of Fe–Si alloys.}
    Results for B2 (red), hcp (blue), fcc (green), and bcc (orange) phases with different compositions at
    (\textbf{a}) 3000~K, (\textbf{b}) 4000~K, (\textbf{c}) 5000~K, and (\textbf{d}) 6000~K. All Gibbs free energy curves were rescaled by removing the linear part from the hcp phase such that $\Delta G = 0$ at both ends. The dashed lines indicate the common tangent between B2 and hcp phases.}
    \label{fig:G}
\end{figure*}

\begin{figure*}[htp]
    \centering
    \includegraphics[width=0.9\linewidth]{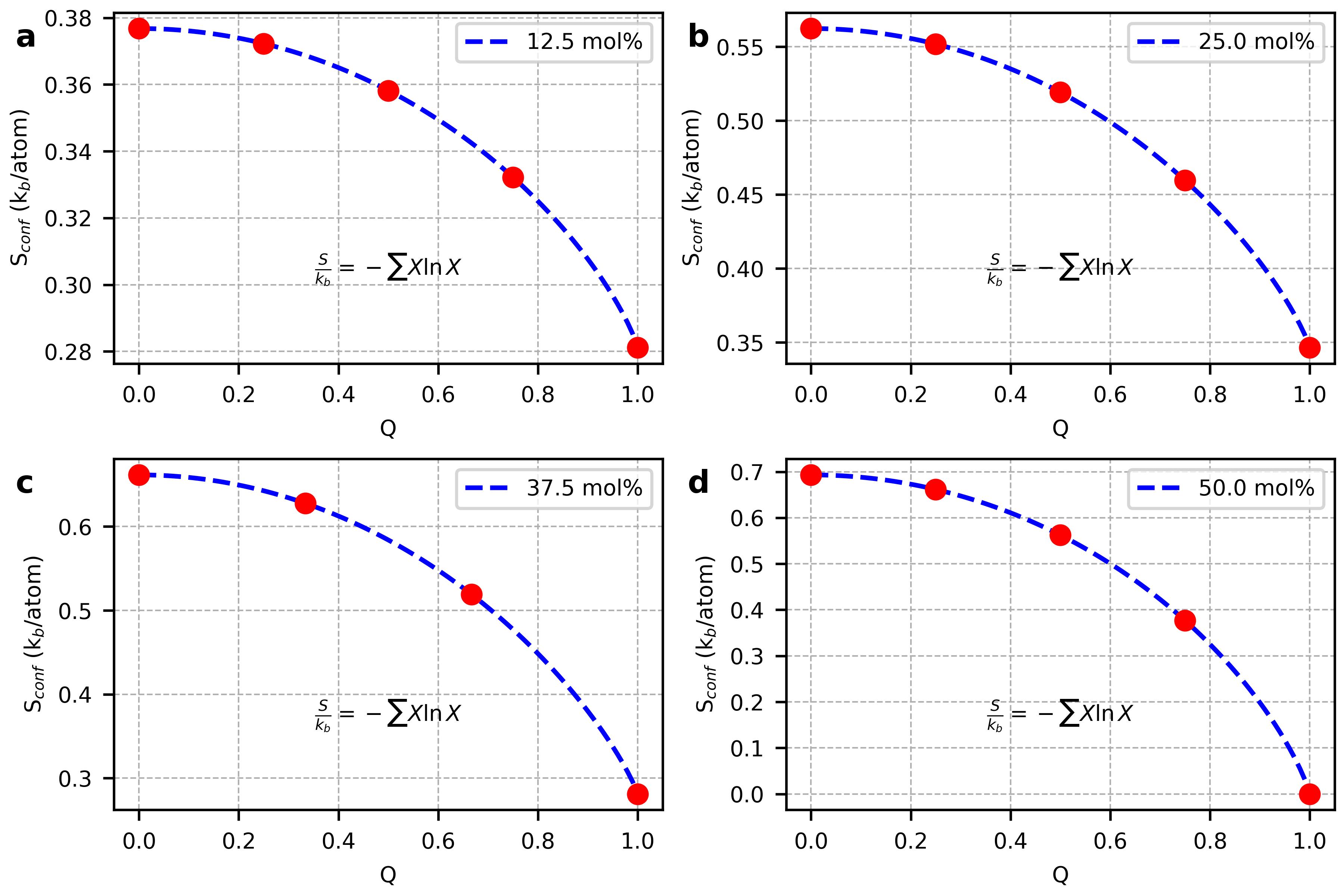}
    \caption{\textbf{Evolution of configurational entropy in the two-site model.}
    The entropy varies with the order parameter $Q$ for Si concentrations of 12.5, 25.0, 37.5, and 50.0~mol\%. $X$ denotes the occupation of Fe and Si atoms at each site.}
    \label{fig:S}
\end{figure*}

\begin{figure*}[htp]
    \centering
    \includegraphics[width=0.9\linewidth]{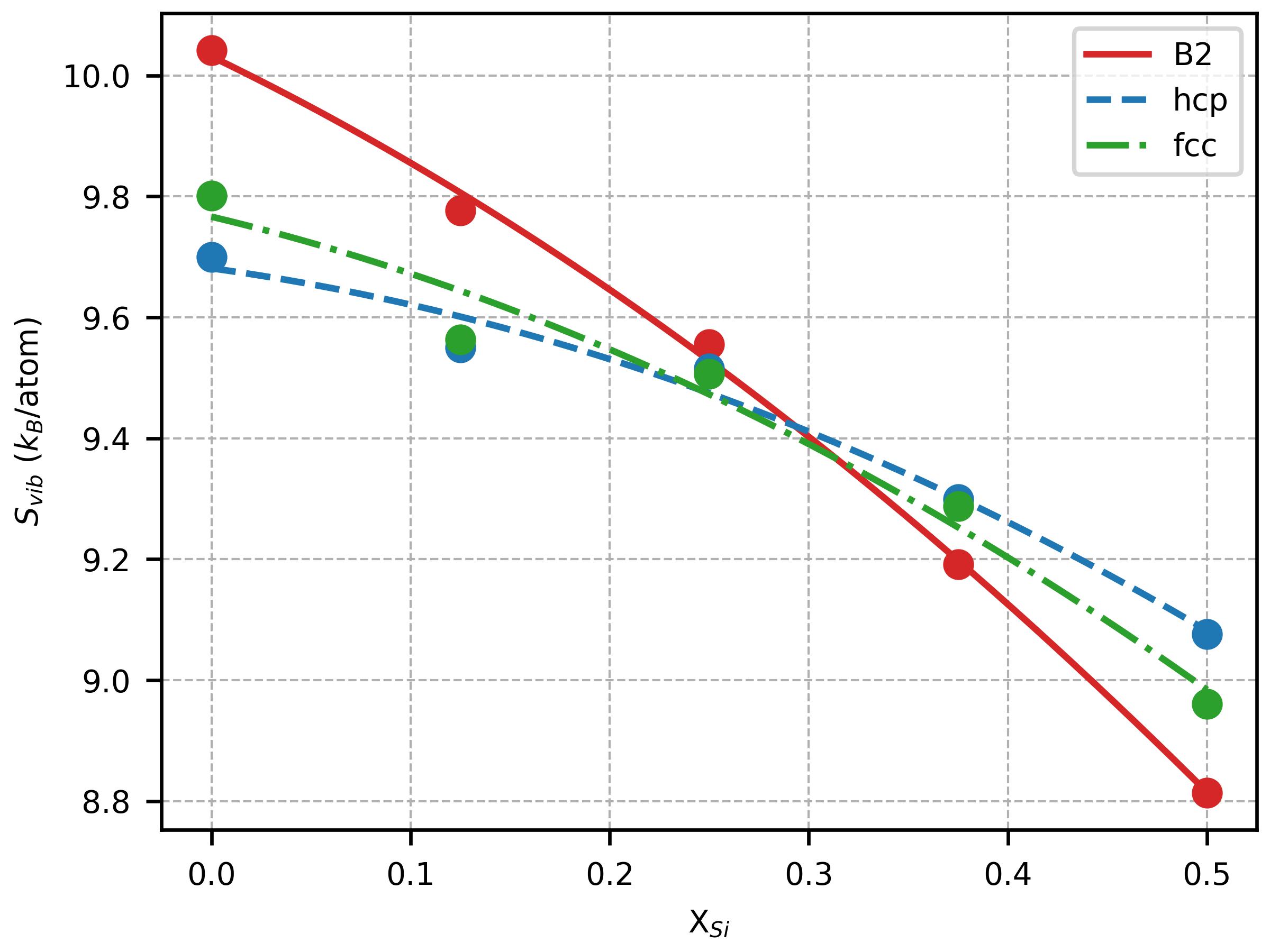}
    \caption{\textbf{Evolution of vibrational entropy with Si content.}
    Results obtained from molecular dynamics simulations at 330~GPa and 6000~K. The vibrational entropy exhibits minor variation among different order parameters ($Q$) for Si contents below 12.5~mol\%.}
    \label{fig:S_vib}
\end{figure*}

\begin{figure*}[htp]
    \centering
    \includegraphics[width=0.8\linewidth]{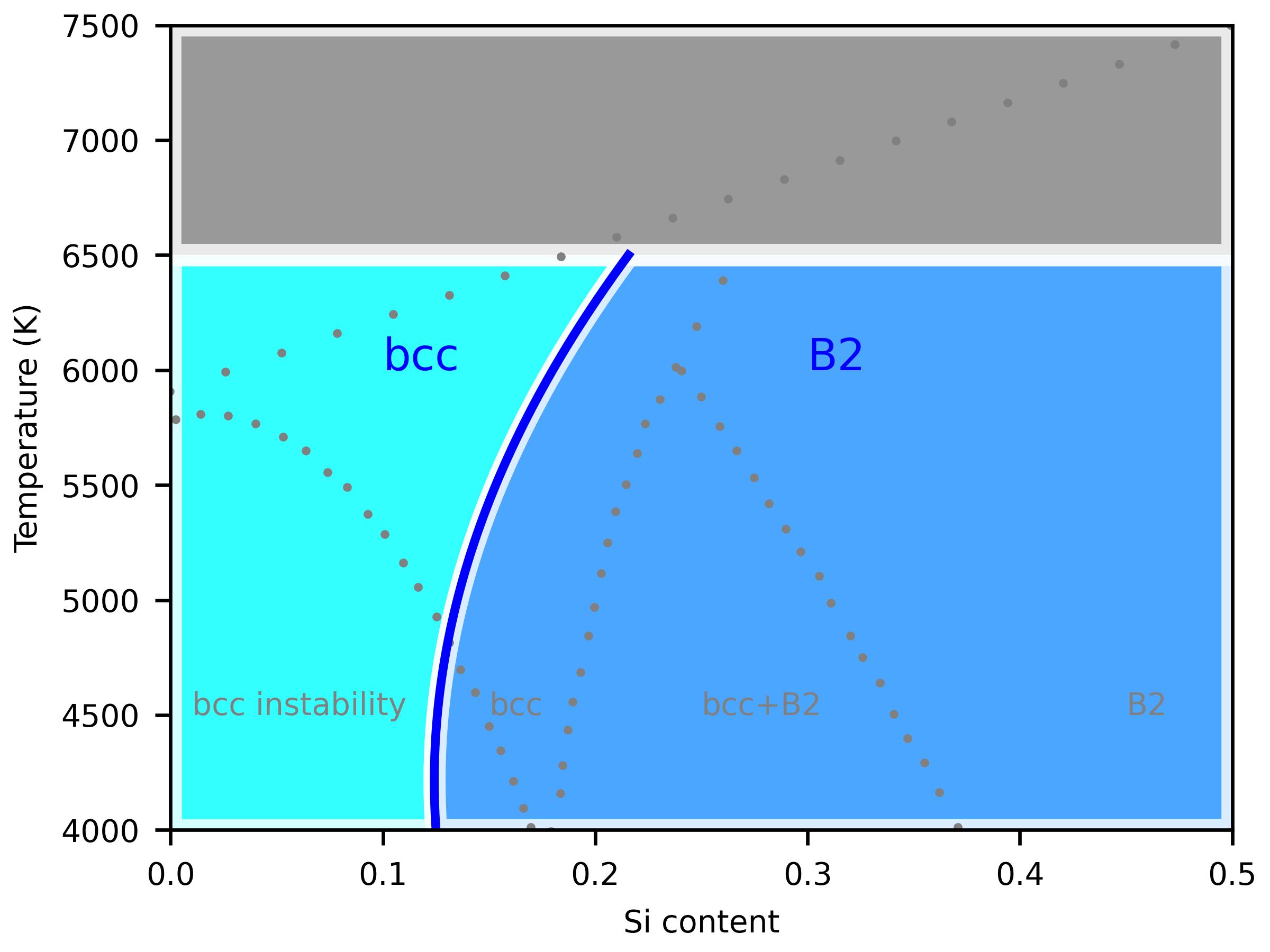}
    \caption{\textbf{Schematic diagram of the Fe--Si system at 330 GPa} restricted to the bcc and B2 phases and showing the lowest free energy single phase at each point. The solid blue line represents the phase boundary between bcc (cyan) and B2 (blue), showing that the bcc phase is stabilized at higher temperatures and lower Si concentrations. For comparison, the phase boundary from Ref.~\cite{liShortrangeOrderStabilizes2025} is shown as dashed gray lines, indicating a coexistence region between the bcc and B2 phases. The shaded grey regions above 6300 K indicate the constraints imposed by melting curves beyond the range of our simulations.}
    \label{fig:B2-phase-diagram}
\end{figure*}

\begin{figure*}[htp]
    \centering
    \includegraphics[width=0.9\linewidth]{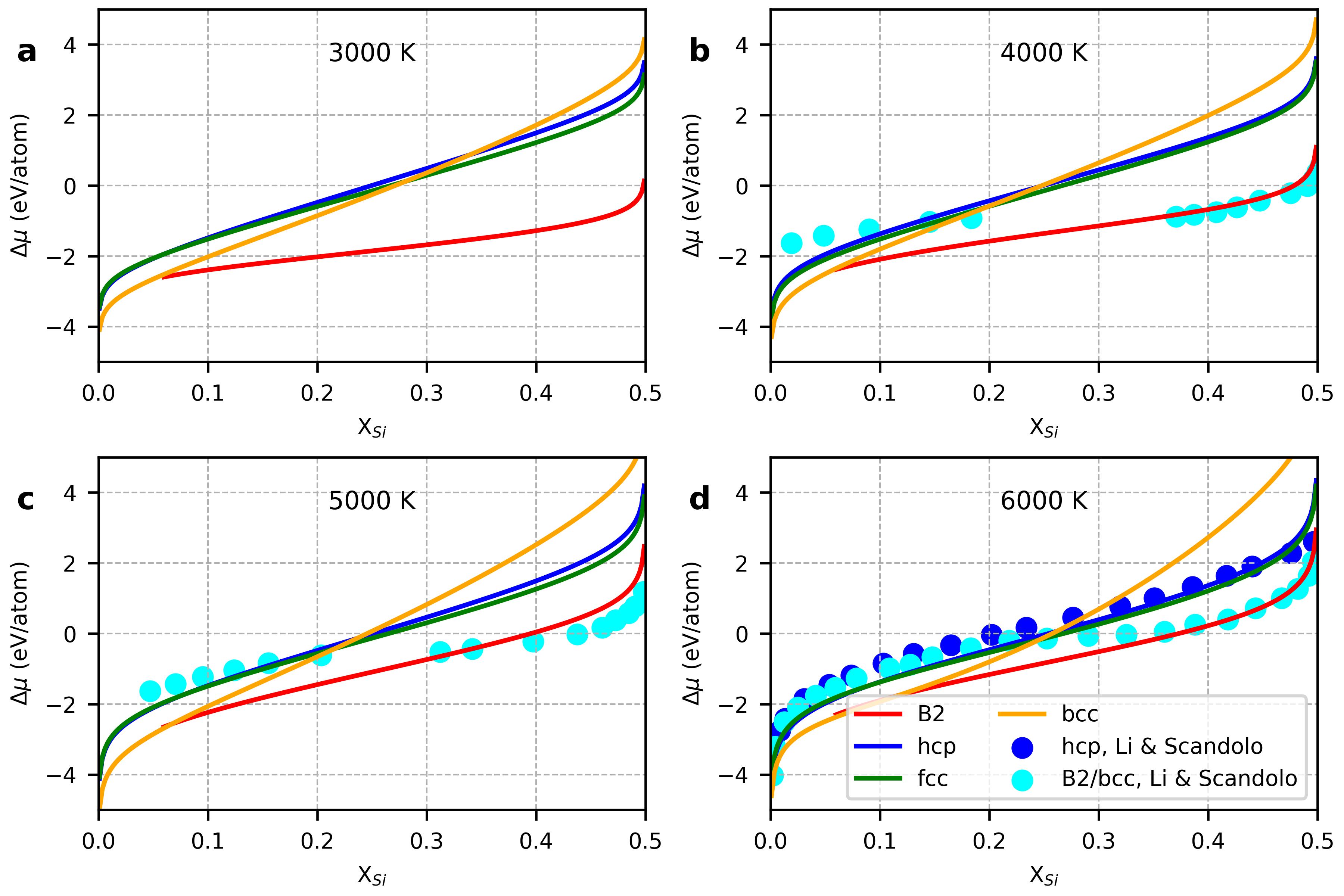}
    \caption{\textbf{Chemical potential difference ($\Delta\mu$)} between Si and Fe as a function of Si concentration, derived from the Gibbs free energy curves of the B2 (red), hcp (blue), fcc (green), and bcc (orange) phases at (\textbf{a})~3000~K, (\textbf{b})~4000~K, (\textbf{c})~5000~K, and (\textbf{d})~6000~K. For comparison, chemical potentials from Li and Scandolo~\cite{liShortrangeOrderStabilizes2025} are shown as reference points for the hcp phase (blue symbols) and the B2/bcc phases (cyan symbols).}
    \label{fig:dG}
\end{figure*}

\begin{figure*}[htp]
    \centering
    \includegraphics[width=0.9\linewidth]{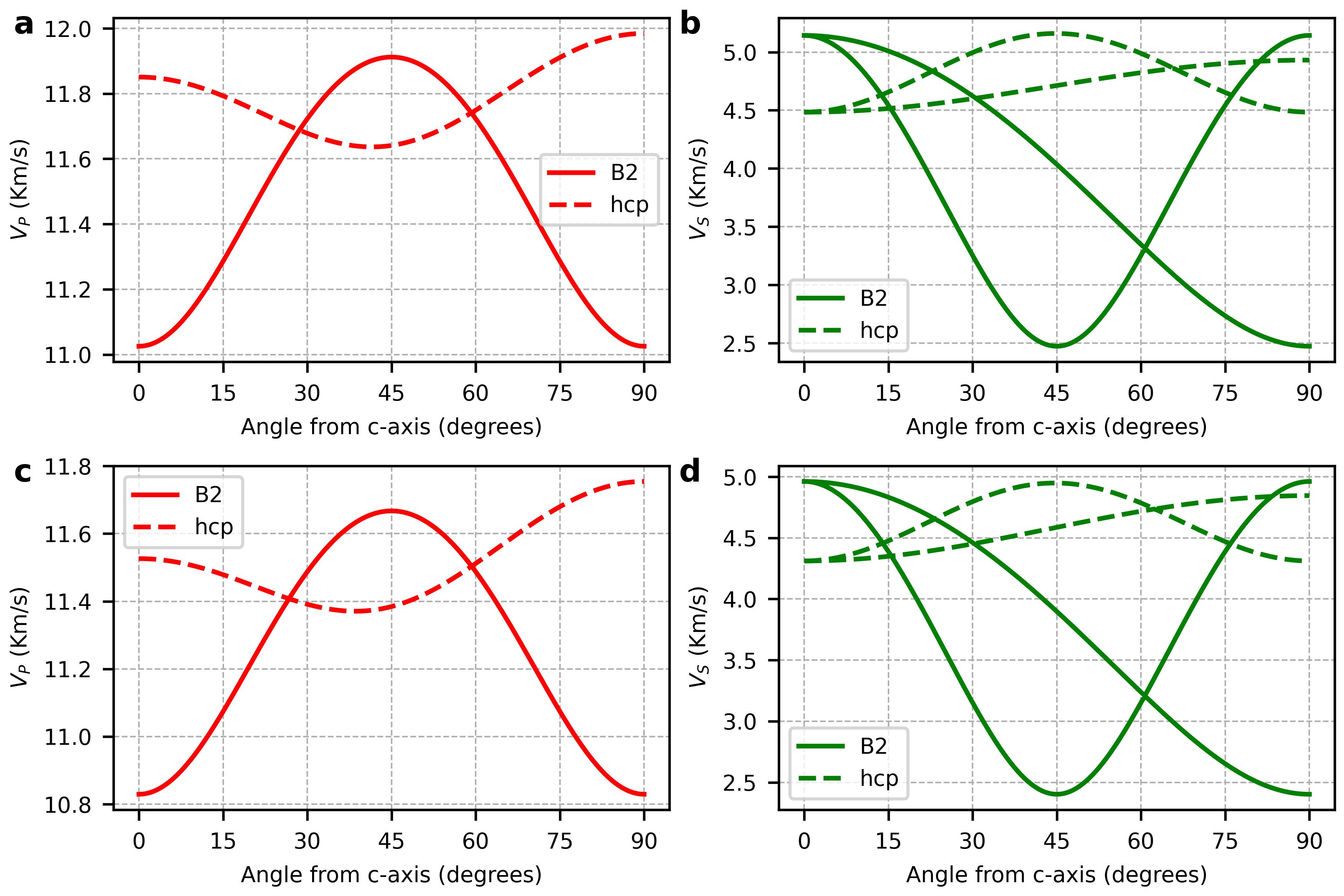}
    \caption{\textbf{Directional dependence of seismic wave velocities.}
    Compressional ($V_P$) and shear ($V_S$) wave velocities of B2 and hcp Fe$_{0.875}$Si$_{0.125}$ alloys as functions of propagation direction relative to the $c$-axis at 5000~K (\textbf{a}, \textbf{b}) and 6000~K (\textbf{c}, \textbf{d}). In the B2 structure, Fe$_{0.875}$Si$_{0.125}$ stabilizes in the disordered ($Q=0$, bcc) phase.}
    \label{fig:axial}
\end{figure*}

\begin{figure*}[htp]
    \centering
    \includegraphics[width=0.8\linewidth]{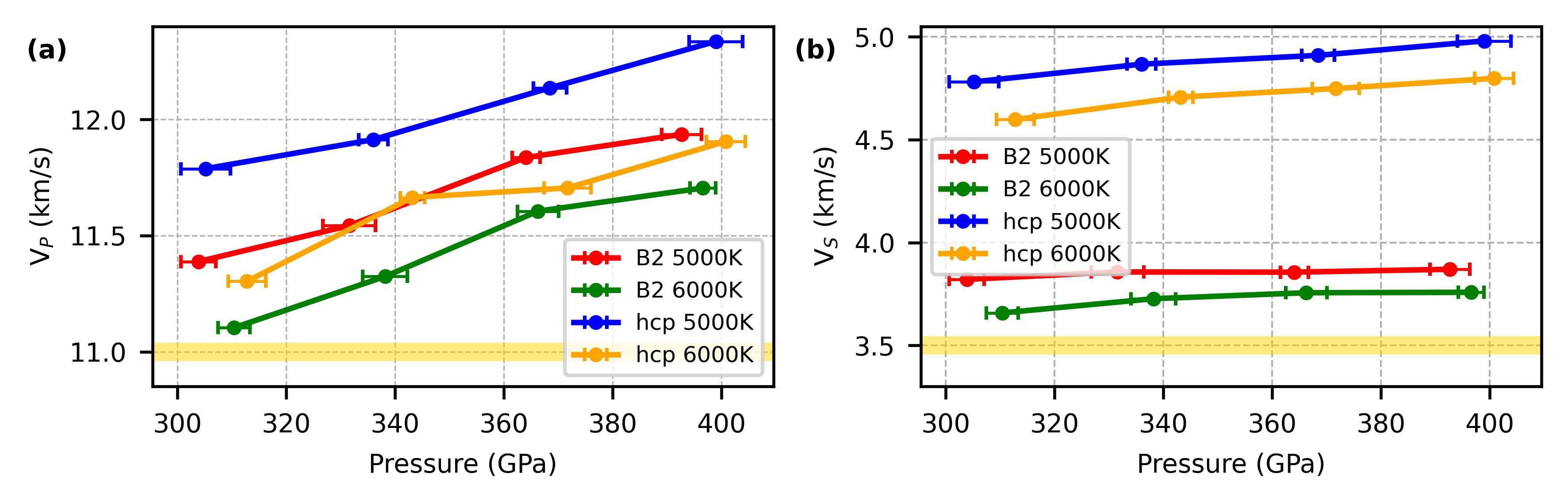}
    \caption{\textbf{Seismic velocities of Fe–Si alloys as a function of pressure.}
    Calculated (\textbf{a}) compressional-wave (\(V_P\)) and (\textbf{b}) shear-wave (\(V_S\)) velocities of Fe$_{0.875}$Si$_{0.125}$ alloys in the B2 and hcp phases at 5000~K and 6000~K over the pressure range 300–400~GPa. The yellow band indicates the average seismic velocities of the Earth's inner core~\cite{dziewonskiPreliminaryReferenceEarth1981}. The error bars indicate statistical pressure errors from MD simulations. Within the cubic structure, Fe$_{0.875}$Si$_{0.125}$ stabilizes in a disordered bcc-like configuration ($Q=0$). Both \(V_P\) and \(V_S\) increase with pressure and decrease with temperature, reflecting compression stiffening and thermal softening, respectively. Compared with the B2 phase, the hcp phase exhibits higher seismic velocities under all conditions, whereas disordered bcc Fe$_{0.875}$Si$_{0.125}$ shows shear-wave velocities closer to the seismological reference values.}

    \label{fig:VPX}
\end{figure*}

\begin{figure*}[htp]
    \centering
    \includegraphics[width=0.9\linewidth]{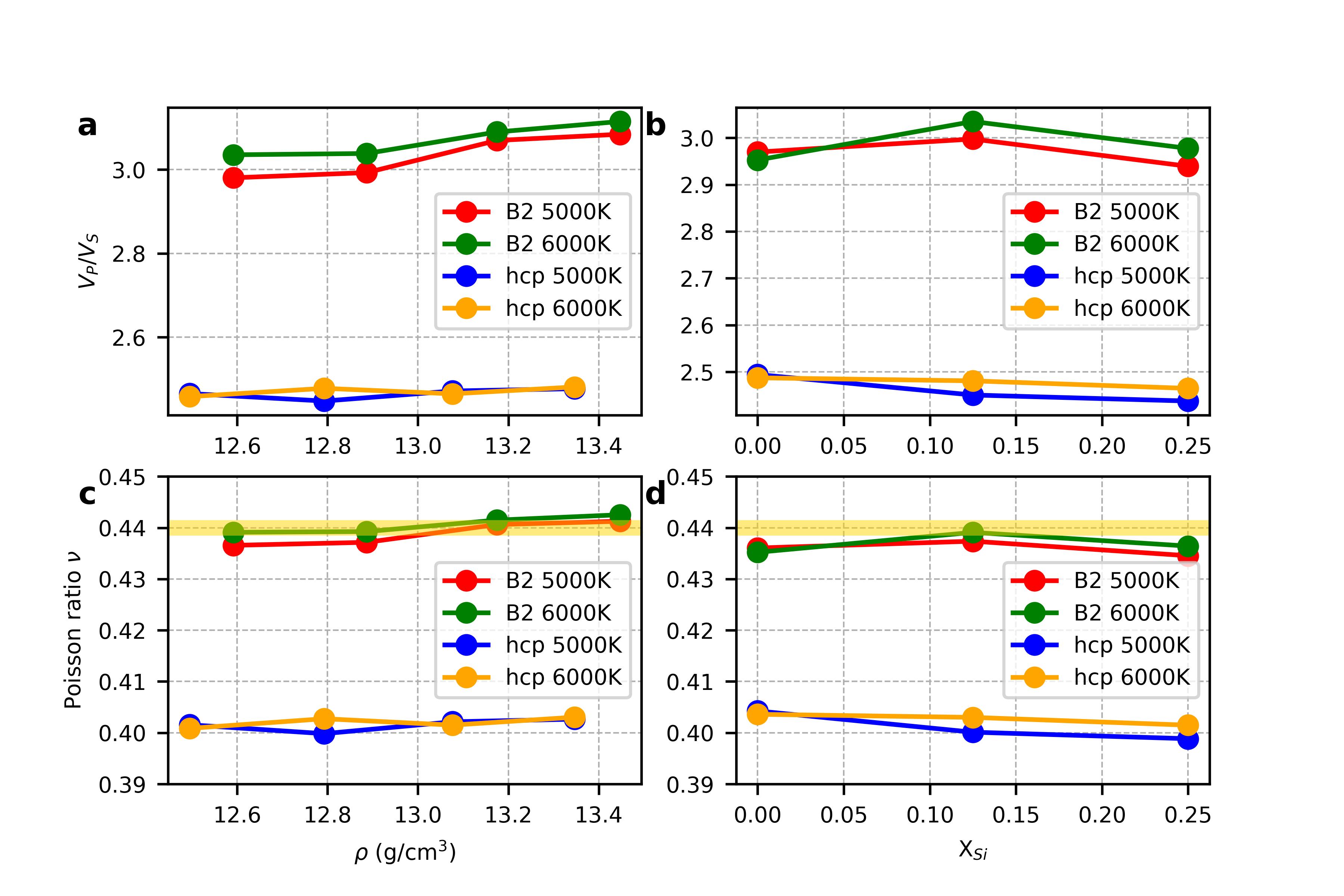}
    \caption{\textbf{Variation of $V_P/V_S$ and Poisson’s ratio $\nu$} of Fe–Si alloys with (\textbf{a, c}) density and (\textbf{b, d}) composition in B2 and hcp phases at 5000~K and 6000~K. The yellow region marks the observed Poisson’s ratio of $\sim$0.44 \cite{dziewonskiPreliminaryReferenceEarth1981}. In the B2 structure, Fe$_{0.875}$Si$_{0.125}$ stabilizes in the disordered ($Q=0$, bcc) phase, while Fe$_{0.75}$Si$_{0.25}$ remains ordered ($Q=1$) at 5000–6000~K. The B2 phase exhibits a significantly higher $V_P/V_S$ ratio than the hcp phase, yielding a Poisson's ratio that closely matches the seismological observation of 0.44.}
    \label{fig:poisson}
\end{figure*}

\begin{figure*}[htp]
    \centering
    \includegraphics[width=0.8\linewidth]{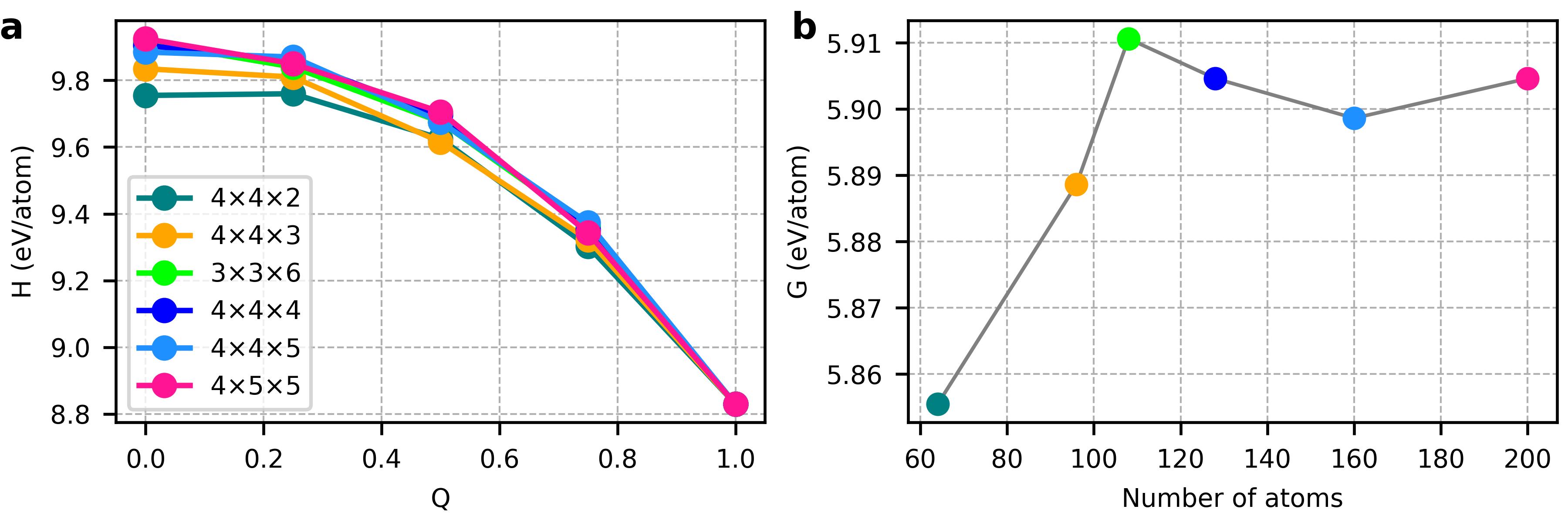}
    \caption{\textbf{Effect of supercell size on SQS simulations.}
    (\textbf{a}) Enthalpy of B2 Fe$_{0.5}$Si$_{0.5}$ at 330~GPa as a function of $Q$, computed using different supercell sizes.
    (\textbf{b}) Gibbs free energy of disordered B2 phase (bcc, $Q=0$) Fe–Si at 6000~K versus supercell atom number.}
    \label{fig:Q_test}
\end{figure*}

\begin{figure*}[htp]
    \centering
    \includegraphics[width=0.9\linewidth]{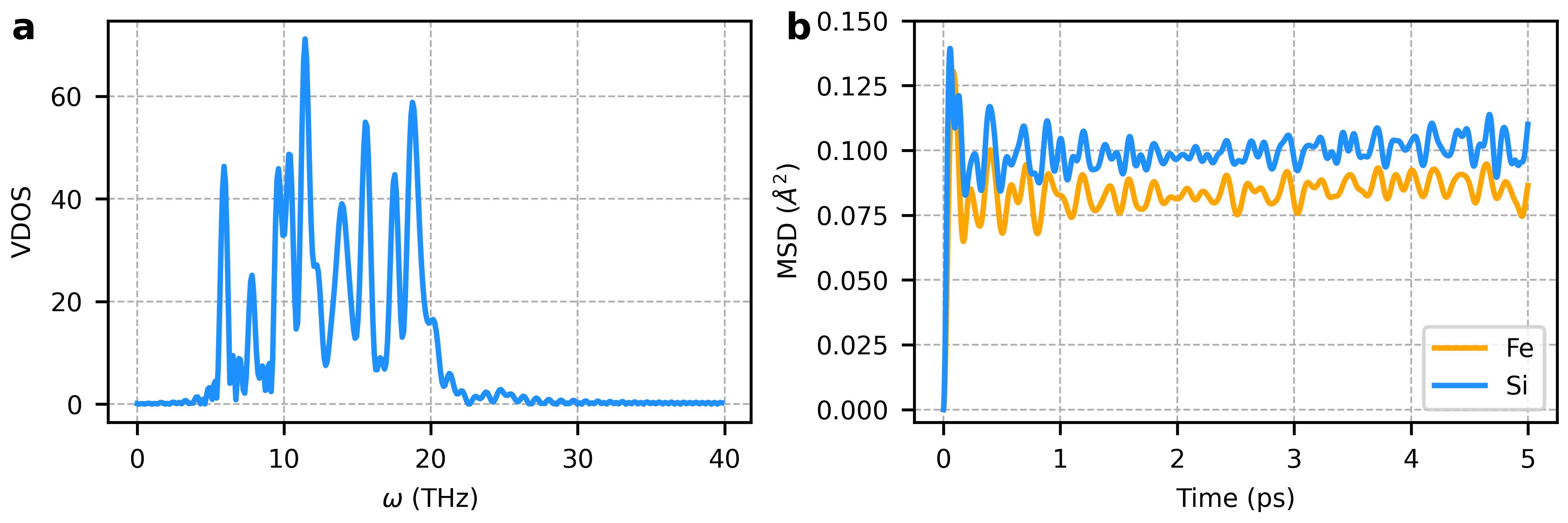}
    \caption{\textbf{Vibrational properties of Fe$_{0.875}$Si$_{0.125}$ alloy at Earth's inner-core conditions.}
    (\textbf{a}) Vibrational density of states (VDOS) and (\textbf{b}) mean-square displacement obtained from first-principles molecular dynamics (FPMD) simulations indicate solid-like atomic behavior without diffusivity.}
    \label{fig:MD}
\end{figure*}

\begin{figure}[htp]
\centering
\includegraphics[width=0.85\linewidth]{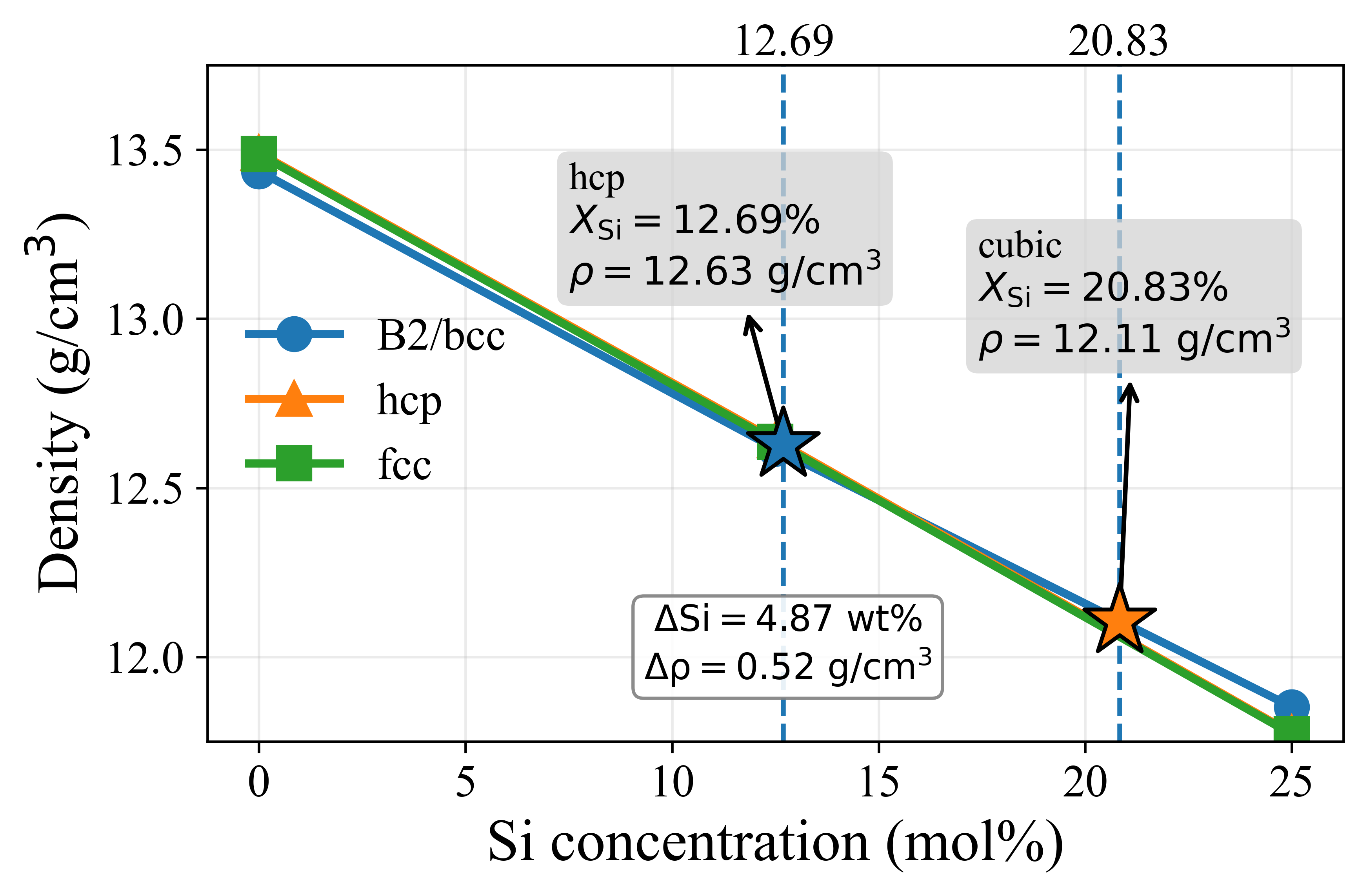}
\caption{
Density variation of Fe--Si phases as a function of silicon concentration at
330~GPa and 6000~K. The B2/bcc, hcp, and fcc phases are shown by blue
circles, orange triangles, and green squares, respectively. For the cubic
phase, the disorder--order parameter $Q$ is labeled next to each data point,
where $Q=0$ and $Q=1$ represent the disordered bcc-like and ordered B2
configurations, respectively. The hcp and fcc phases are included as
reference close-packed structures. Dashed vertical lines indicate the
equilibrium hcp and cubic compositions obtained from the common-tangent
construction, $x_{\mathrm{Si}}=12.69$~mol\% and
$x_{\mathrm{Si}}=20.83$~mol\%, respectively. The corresponding densities
are approximately 12.63 and 12.11~g\,cm$^{-3}$.
The two phases therefore differ by
$\Delta\mathrm{Si}=4.87$~wt\% in composition and
$\Delta\rho=0.52$~g\,cm$^{-3}$ in density, with the Si-richer cubic phase
being less dense than the hcp phase. This density contrast provides an
additional constraint on possible phase distributions during inner-core
evolution, although their actual spatial distribution also depends on
compositional partitioning, pressure, temperature, phase fractions, and
crystallization kinetics.
}
\label{fig:density_coexistence}
\end{figure}

\begin{enumerate}
    \item Fig.~\ref{fig:atoms} illustrates the two-site model in the B2, hcp, and fcc phases.
    \item Fig.~\ref{fig:H} shows the enthalpy as a function of the order parameter $Q$ at 330~GPa.
    \item Fig.~\ref{fig:xrd} presents the powder diffraction patterns of the bcc and B2 phases of the Fe–Si alloy.
    \item Fig.~\ref{fig:G} shows the Gibbs free energy $\Delta G$ of the Fe–Si alloy.
    \item Fig.~\ref{fig:S} illustrates the evolution of the configurational entropy in the two-site model.
    \item Fig.~\ref{fig:S_vib} shows the evolution of vibrational entropy with composition at 6000~K from molecular dynamics simulations.
    \item Fig.~\ref{fig:dG} shows the chemical potential $\Delta\mu$ of the Fe-Si alloy.
    \item Fig.~\ref{fig:B2-phase-diagram} illustrates the thermodynamic competition between the B2 and bcc phases across varying temperatures and compositions at 330 GPa.
    conditions, indicating solid-like atomic behavior.
    \item Fig.~\ref{fig:axial} presents the compressional wave velocity ($V_P$) and shear wave velocity ($V_S$) as functions of propagation direction relative to the $c$-axis.
    \item Fig.~\ref{fig:VPX} shows the pressure dependence of seismic velocities.
    \item Fig.~\ref{fig:poisson} shows the density- and composition-dependent Poisson’s ratio.
    \item Fig.~\ref{fig:Q_test} shows the effect of supercell size on the SQS simulations.
    \item Fig.~\ref{fig:MD} shows the vibrational density of states and mean-square displacement of Fe$_{0.875}$Si$_{0.125}$ alloy under Earth's inner-core 
    \item  Fig.~\ref{fig:density_coexistence} shows the density of different phases with respect to Si content at 330 GPa and 6000 K.
    \item Table~\ref{table:atom} lists the Fe and Si concentrations at each site in the two-site model.
    \item Table~\ref{table:cij} lists the elastic constants of the disordered-phase Fe$_{0.875}$Si$_{0.125}$ alloy under Earth's inner-core conditions.
\end{enumerate}

\newpage

\newpage

\newpage

\newpage

\newpage

\newpage


\newpage

\newpage

\newpage

\newpage

\newpage

\newpage

\newpage

\newpage
\begin{table*}[htp]
  \centering
  \caption{Composition, order parameter $Q$, and Fe/Si concentrations at each site in the two-site model.}
  \begin{tabular}{cccccc}
    \hline\hline
    \multirow{2}{*}{Composition} & \multirow{2}{*}{$Q$} & \multicolumn{2}{c}{Site 1} & \multicolumn{2}{c}{Site 2} \\
    \cline{3-6}
    & & Fe$_1$ & Si$_1$ & Fe$_2$ & Si$_2$ \\
    \hline
    Fe$_{1.000}$Si$_{0.000}$ & 1.000 & 1.00000 & 0.00000 & 1.00000 & 0.00000 \\
    \hline
    \multirow{5}{*}{Fe$_{0.875}$Si$_{0.125}$} 
      & 0.000 & 0.87500 & 0.12500 & 0.87500 & 0.12500 \\
      & 0.250 & 0.90625 & 0.09375 & 0.84375 & 0.15625 \\
      & 0.500 & 0.93750 & 0.06250 & 0.81250 & 0.18750 \\
      & 0.750 & 0.96875 & 0.03125 & 0.78125 & 0.21875 \\
      & 1.000 & 1.00000 & 0.00000 & 0.75000 & 0.25000 \\
    \hline
    \multirow{5}{*}{Fe$_{0.750}$Si$_{0.250}$}
      & 0.000 & 0.75000 & 0.25000 & 0.75000 & 0.25000 \\
      & 0.250 & 0.81250 & 0.18750 & 0.68750 & 0.31250 \\
      & 0.500 & 0.87500 & 0.12500 & 0.62500 & 0.37500 \\
      & 0.750 & 0.93750 & 0.06250 & 0.56250 & 0.43750 \\
      & 1.000 & 1.00000 & 0.00000 & 0.50000 & 0.50000 \\
    \hline
    \multirow{4}{*}{Fe$_{0.625}$Si$_{0.375}$}
      & 0.000 & 0.62500 & 0.37500 & 0.62500 & 0.37500 \\
      & 0.333 & 0.75000 & 0.25000 & 0.50000 & 0.50000 \\
      & 0.667 & 0.87500 & 0.12500 & 0.37500 & 0.62500 \\
      & 1.000 & 1.00000 & 0.00000 & 0.25000 & 0.75000 \\
    \hline
    \multirow{5}{*}{Fe$_{0.500}$Si$_{0.500}$}
      & 0.000 & 0.50000 & 0.50000 & 0.50000 & 0.50000 \\
      & 0.250 & 0.62500 & 0.37500 & 0.37500 & 0.62500 \\
      & 0.500 & 0.75000 & 0.25000 & 0.25000 & 0.75000 \\
      & 0.750 & 0.87500 & 0.12500 & 0.12500 & 0.87500 \\
      & 1.000 & 1.00000 & 0.00000 & 0.00000 & 1.00000 \\
    \hline\hline
  \end{tabular}
  \label{table:atom}
\end{table*}

\newpage
\begin{table}[htp]
  \centering
  \caption{Densities ($\rho$), temperature (T), pressure (P), elastic constants (C$_{ij}$), bulk modulus (B), Hill averaged shear modulus (G), seismic velocities (V$_P$, V$_S$) and Chung–Buessem anisotropy (A$_{CB}$) of pure Fe and disordered Fe$_{0.875}$Si$_{0.125}$ alloy in B2 and hcp phases at 5000 K, 6000 K, and 6500 K. In the B2 structure, Fe$_{0.875}$Si$_{0.125}$ stabilizes in the disordered ($Q=0$, bcc-like) phase, while Fe$_{0.75}$Si$_{0.25}$ remains ordered ($Q=1$) at 5000–6000~K.}
  \label{table:cij}
  \resizebox{\textwidth}{!}{%
  \begin{tabular}{c c c c c c c c c c c c c c c}
    \hline\hline
    phase & comp. & $\rho$ & $T$ & $P$
    & $C_{11}$ & $C_{12}$ & $C_{44}$ & $C_{13}$ & $C_{33}$
    & $B$ & $G$ & $V_P$ & $V_S$ & $A_{CB}$ \\
    & & (g/cm$^3$) & (K) & (GPa)
    & (GPa) & (GPa) & (GPa) & (GPa) & (GPa)
    & (GPa) & (GPa) & (km/s) & (km/s) & (\%) \\
    \hline
    bcc & pure Fe & 13.55 & 5000 & 333 & 1427 & 1280 & 316 & -- & -- & 1329 & 178 & 10.75 & 3.62 & 23.4 \\
    bcc & pure Fe & 13.55 & 6000 & 337 & 1397 & 1236 & 292 & -- & -- & 1290 & 175 & 10.60 & 3.59 & 18.7 \\
    hcp & pure Fe & 13.72 & 5000 & 338 & 1719 & 1178 & 248 & 1094 & 1684 & 1316 & 269 & 11.05 & 4.43 & 0.4 \\
    hcp & pure Fe & 13.72 & 6000 & 346 & 1642 & 1126 & 243 & 1050 & 1614 & 1260 & 260 & 10.82 & 4.35 & 0.3 \\
    \hline
    bcc & Fe$_{0.875}$Si$_{0.125}$ & 12.79 & 5000 & 332 & 1555 & 1398 & 337 & -- & -- & 1450 & 190 & 11.54 & 3.85 & 23.3 \\
    bcc & Fe$_{0.875}$Si$_{0.125}$ & 12.79 & 6000 & 338 & 1501 & 1352 & 315 & -- & -- & 1402 & 178 & 11.32 & 3.73 & 22.9 \\
    bcc & Fe$_{0.875}$Si$_{0.125}$ & 12.79 & 6500 & 345 & 1452 & 1311 & 295 & -- & -- & 1358 & 168 & 11.12 & 3.62 & 22.4 \\
    hcp & Fe$_{0.875}$Si$_{0.125}$ & 12.89 & 5000 & 336 & 1891 & 1249 & 266 & 1170 & 1847 & 1422 & 304 & 11.91 & 4.86 & 0.7 \\
    hcp & Fe$_{0.875}$Si$_{0.125}$ & 12.89 & 6000 & 343 & 1814 & 1201 & 250 & 1147 & 1746 & 1373 & 285 & 11.66 & 4.70 & 0.6 \\
    hcp & Fe$_{0.875}$Si$_{0.125}$ & 12.89 & 6500 & 347 & 1785 & 1170 & 240 & 1092 & 1681 & 1327 & 281 & 11.49 & 4.67 & 0.9 \\
    \hline
    B2 & Fe$_{0.75}$Si$_{0.25}$ & 11.96 & 5000 & 335 & 1589 & 1420 & 357 & -- & -- & 1476 & 202 & 12.08 & 4.11 & 22.8 \\
    B2 & Fe$_{0.75}$Si$_{0.25}$ & 11.96 & 6000 & 343 & 1532 & 1374 & 337 & -- & -- & 1427 & 189 & 11.85 & 3.98 & 23.2 \\
    hcp & Fe$_{0.75}$Si$_{0.25}$ & 12.10 & 5000 & 339 & 1947 & 1346 & 297 & 1180 & 1964 & 1473 & 320 & 12.53 & 5.14 & 0.9 \\
    hcp & Fe$_{0.75}$Si$_{0.25}$ & 12.10 & 6000 & 346 & 1857 & 1298 & 282 & 1147 & 1868 & 1417 & 299 & 12.25 & 4.97 & 0.8 \\

    \hline\hline
  \end{tabular}%
  }
\end{table}

\end{document}